\documentstyle[emlines2]{article}

\def\hybrid{\topmargin 0pt      \oddsidemargin 0pt
        \headheight 0pt \headsep 0pt
        \textwidth 6.25in       
        \textheight 9.5in       
        \marginparwidth 0.0in
        \parskip 5pt plus 1pt   \jot = 1.5ex}
\catcode`\@=11
\def\marginnote#1{}

\newcount\hour
\newcount\minute
\newtoks\amorpm
\hour=\time\divide\hour by60
\minute=\time{\multiply\hour by60 \global\advance\minute by-\hour}
\edef\standardtime{{\ifnum\hour<12 \global\amorpm={am}%
        \else\global\amorpm={pm}\advance\hour by-12 \fi
        \ifnum\hour=0 \hour=12 \fi
        \number\hour:\ifnum\minute<10 0\fi\number\minute\the\amorpm}}
\edef\militarytime{\number\hour:\ifnum\minute<10 0\fi\number\minute}

\def\draftlabel#1{{\@bsphack\if@filesw {\let\thepage\relax
   \xdef\@gtempa{\write\@auxout{\string
      \newlabel{#1}{{\@currentlabel}{\thepage}}}}}\@gtempa
   \if@nobreak \ifvmode\nobreak\fi\fi\fi\@esphack}
        \gdef\@eqnlabel{#1}}
\def\@eqnlabel{}
\def\@vacuum{}
\def\draftmarginnote#1{\marginpar{\raggedright\scriptsize\tt#1}}

\def\draftlabel#1{{\@bsphack\if@filesw {\let\thepage\relax
   \xdef\@gtempa{\write\@auxout{\string
      \newlabel{#1}{{\@currentlabel}{\thepage}}}}}\@gtempa
   \if@nobreak \ifvmode\nobreak\fi\fi\fi\@esphack}
        \gdef\@eqnlabel{#1}}
\def\@eqnlabel{}
\def\@vacuum{}
\def\draftmarginnote#1{\marginpar{\raggedright\scriptsize\tt#1}}

\def\draft{\oddsidemargin -.5truein
        \def\@oddfoot{\sl preliminary draft \hfil
        \rm\thepage\hfil\sl\today\quad\militarytime}
        \let\@evenfoot\@oddfoot \overfullrule 3pt
        \let\label=\draftlabel
        \let\marginnote=\draftmarginnote
   \def\@eqnnum{(\theequation)\rlap{\kern\marginparsep\tt\@eqnlabel}%
\global\let\@eqnlabel\@vacuum}  }

\def\numberbysection{\@addtoreset{equation}{section}
        \def\theequation{\thesection.\arabic{equation}}}

\def\underline#1{\relax\ifmmode\@@underline#1\else
        $\@@underline{\hbox{#1}}$\relax\fi}

\def\titlepage{\@restonecolfalse\if@twocolumn\@restonecoltrue\onecolumn
     \else \newpage \fi \thispagestyle{empty}\c@page\z@
        \def\thefootnote{\fnsymbol{footnote}} }

\def\endtitlepage{\if@restonecol\twocolumn \else  \fi
        \def\thefootnote{\arabic{footnote}}
        \setcounter{footnote}{0}}  
\relax

\numberbysection
\hybrid

\def\beq{\begin{equation}}
\def\eeq{\end{equation}}
\def\p{\partial}

\newtheorem{th}{Theorem}[section]

\newtheorem{lem}{Lemma}[section]

\newdimen\Squaresize \Squaresize=30pt
\newdimen\Thickness \Thickness=0.5pt

\def\Square#1{\hbox{\vrule width \Thickness
   \vbox to \Squaresize{\hrule height \Thickness\vss
      \hbox to \Squaresize{\hss#1\hss}
   \vss\hrule height\Thickness}
\unskip\vrule width \Thickness}    
\kern-\Thickness}                  

\def\Vsquare#1{\vbox{\Square{$#1$}}\kern-\Thickness}

\def\young#1{
\vbox{\smallskip\offinterlineskip
\halign{&\Vsquare{##}\cr #1}}}

\begin{document}

\begin{titlepage}

\title{Discrete Hirota's equation in quantum integrable models}

\author{A. Zabrodin
\thanks{Joint Institute of Chemical Physics, Kosygina str. 4, 117334,
Moscow, Russia and ITEP, 117259, Moscow, Russia}}

\maketitle

\begin{abstract}

The recent progress in revealing classical integrable structures
in quantum models solved by Bethe ansatz is reviewed. Fusion relations
for eigenvalues of quantum transfer matrices can be written in the form
of classical Hirota's bilinear difference equation. This equation
is also known as the completely discretized version of the 2D Toda
lattice. We explain how one obtains the specific quantum results
by solving the classical equation. The auxiliary linear problem for
the Hirota equation is shown to generalize Baxter's $T$-$Q$
relation.

\end{abstract}

\vfill

\end{titlepage}

\section{Introduction}

Acquiring some experience in classical and quantum
integrable systems, one would probably say that these
theories hardly
have something to do with each other. They do look like different
branches of mathematical physics with their own methods, notions and
traditions. The correspondence is believed to exist but practically it
does not seem to uncover itself to a satisfactory extent.

Actually, the very questions used to be
asked in the classical and quantum cases usually did not have
common points. So there is no surprize that
main ingredients of the theories practically do not
intersect. For instance, the inverse
scattering method has a very different meaning in the two theories
to say nothing about such specific tools as the finite-gap
integration technique (from the classical side) or the Bethe ansatz
(from the quantum side).

At the same time we would like to believe that
such a strong property as integrability may add
something important to the correspondence principle.
At present it is gradually realized that there should
exist a structure common for classical and quantum integrability, or
rather that classical and quantum cases should be the two faces
of this structure.

Having all this in mind, let us give an outline of a more technical
story presented here. These notes review
the recent progress
(see the first paper in ref.\,\cite{KLWZ}) in uncovering classical
integrable structures in quantum models.
More attention is paid to some questions which remained obscure or were
not very clearly written in \cite{KLWZ}.
In particular, we try to follow,
at least schematically, the
whole long way from the simplest quantum $R$-matrix to the
classical discrete soliton equations.
We hope that in the future the story may be
considerably shortened and simplified.

In Sect.\,2 we review basic elements of the fusion procedure on the
example of the simplest solution to the Yang-Baxter equation with
rational dependence on the spectral parameter. The consisetency of
the fusion procedure relies on the Yang-Baxter equation for the
$R$-matrix. As a result, one obtains a family of quantum
transfer matrices (generating functions for commuting
hamiltonians) acting in one and the same quantum space. They depend
on some (spectral) parameters and commute for all values of these
parameters.

These transfer matrices are functionally dependent. They satisfy a
number of functional relations (called {\it fusion relations}) which are
analogues of relations for characters of linear groups but have a
more complicated structure. Remarkably enough, this structure is
well known in the classical soliton theory in a very different
context. It is an integrable discretization of the 2D Toda
lattice. In the bilinear form it is known as Hirota's bilinear
difference equation (HBDE). This fact suggests an intriguing link
between quantum and classical integrable systems lying much
deeper than any kind of a naive "classical limit". Let us stress
that it is the quantum integrability (i.e. commutativity of the
multiparameter family of quantum operators) that allows one to
reduce operator relations to classical equations for eigenvalues.
In this approach, different quantum states correspond to different
solutions of the classical equation satisfying certain boundary
and analytic conditions.

Sect.\,3 is a brief summary of main facts about HBDE.
From the classical viewpoint, HBDE plays the role of a master equation
for the majority of known continuous and discrete soliton equatuions.
R.Hirota has shown that HBDE unifies various types of them:
the Korteweg-de Vries (KdV) equation,
Kadomtsev-Petviashvili (KP) equation,
two-dimensional Toda lattice (2DTL), the sine-Gordon (SG)
equation, etc as well as their discrete analogues can be obtained from
it by different reductions, specifications of parameters and
continuum limits.
Furthermore, HBDE itself has been shown to
possess soliton solutions and B\"acklund transformations.
The key elements of the theory are discretized Zakharov-Shabat
representation (the zero curvature condition) and the corresponding
auxiliary linear problems.

In Sect.\,4, we show how the specific quantum notions and methods
(such as the transfer matrix, Baxter's $Q$-operators, the nested
Bethe ansatz, etc) are translated into the purely classical language.
In other words, solutions of discrete classical nonlinear equations
appear in the Bethe ansatz form, a surprizing fact that still waits for
a deeper understanding.

\section{Quantum fusion relations}

\subsubsection*{$R$-matrix}

The fundamental rational $R$-matrix acting in
${\bf C}^{k}\times {\bf C}^{k}$
has the form
\beq
R(u)=u+2{\cal P}\,,
\label{A1}
\eeq
where ${\cal P}$ is the permutation operator,
${\cal P}(x\otimes y)=y\otimes x$, and $u$ is the spectral
parameter. It is convenient to represent $R(u)$ as $k\times k$
matrix in the first space
${\bf C}^{k}$
(called the {\it auxiliary space}) with non-commutative entries
which in their turn are $k\times k$ matrices acting in the second
space
${\bf C}^{k}$ (called the {\it quantum space}):
\beq
R(u)=uI+2\sum _{i,j=1}^{k}E_{ij}\otimes E_{ji}\,.
\label{A2}
\eeq
Here $I$ is the unit matrix and $E_{ij}$ is the $k\times k$
matrix such that $(E_{ij})_{mn}=\delta _{im}\delta _{jn}$.

This $R$-matrix satisfies the Yang-Baxter equation
\beq
R_{12}(u_1 -u_2 )
R_{13}(u_1 -u_3 )
R_{23}(u_2 -u_3 )=
R_{23}(u_2 -u_3 )
R_{13}(u_1 -u_3 )
R_{12}(u_1 -u_2 )\,.
\label{YB}
\eeq
Both sides are operators in
${\bf C}^{k}\times {\bf C}^{k}
\times {\bf C}^{k}$. We use the standard notation:
$R_{12}(u)$ acts as $R(u)$ in the tensor product of the first
two spaces and as identity in the third one, similarly for
$R_{13}$, $R_{23}$.
Graphically, eq.\,(\ref{YB}) can be written as follows:

\vspace{5mm}

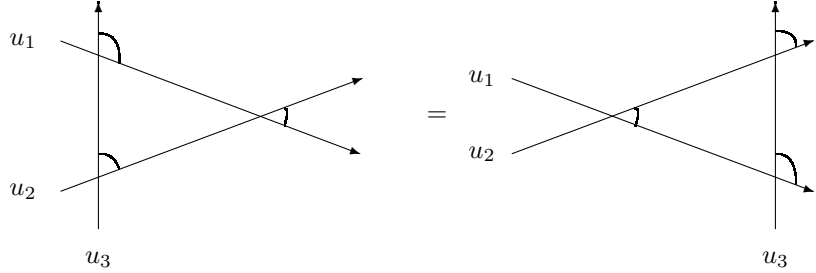
\special{em:linewidth 0.4pt}
\unitlength 1.00mm
\linethickness{0.4pt}
\begin{center}
\begin{picture}(105.00,34.00)
\put(10.00,34.00){\vector(0,1){0.2}}
\emline{10.00}{4.00}{1}{10.00}{34.00}{2}
\put(45.00,24.00){\vector(3,1){0.2}}
\emline{5.00}{9.00}{3}{45.00}{24.00}{4}
\put(44.67,14.00){\vector(3,-1){0.2}}
\emline{5.00}{29.00}{5}{44.67}{14.00}{6}
\put(105.00,29.00){\vector(3,1){0.2}}
\emline{65.00}{14.00}{7}{105.00}{29.00}{8}
\put(100.00,34.00){\vector(0,1){0.2}}
\emline{100.00}{4.00}{9}{100.00}{34.00}{10}
\put(105.00,9.00){\vector(3,-1){0.2}}
\emline{65.00}{24.00}{11}{105.00}{9.00}{12}
\put(55.00,19.00){\makebox(0,0)[cc]{$=$}}
\put(0.00,29.00){\makebox(0,0)[cc]{$u_1$}}
\put(0.00,9.00){\makebox(0,0)[cc]{$u_2$}}
\put(61.00,24.00){\makebox(0,0)[cc]{$u_1$}}
\put(61.00,14.00){\makebox(0,0)[cc]{$u_2$}}
\put(10.00,0.00){\makebox(0,0)[cc]{$u_3$}}
\put(100.00,0.00){\makebox(0,0)[cc]{$u_3$}}
\qbezier(10.00,30.00)(13.00,30.00)(12.67,26.00)
\qbezier(10.00,14.00)(12.00,14.00)(12.67,12.00)
\qbezier(34.67,20.00)(35.33,20.00)(34.67,17.67)
\qbezier(100.00,14.00)(102.67,14.00)(102.67,10.00)
\qbezier(100.00,30.33)(102.67,30.33)(102.67,28.33)
\qbezier(81.33,20.00)(82.00,20.00)(81.33,17.67)
\end{picture}
\end{center}
\vspace{5mm}

\noindent
which means that the line $u_3$ can be moved over the
intersection point of the other two lines.

\subsubsection*{${\cal T}$-matrix}

The quantum monodromy matrix
(${\cal T}$-matrix) is the following product of $R$-matrices
in the auxiliary space $V_0 = {\bf C}^{k}$:
\beq
{\cal T}(u)=R_{0N}(u-y_{N})\ldots
R_{02}(u-y_{2})
R_{01}(u-y_{1})\,.
\label{A3}
\eeq
This is a $k\times k$ matrix with operator entries. They act in the
quantum space $\otimes _{i=1}^{N}V_{i}$, $V_{i}={\bf C}^{k}$.
Matrix elements of $R_{0i}$ and $R_{0j}$ for $i\neq j$ commute
with each other because they operate in different components of the
tensor product. The parameters $y_i$ are arbitrary; sometimes
they are called {\it rapidities}.

The following basic relation is a direct corollary of eq.\,(\ref{YB}):
\beq
R_{12}(u-v){\cal T}_{13}(u)
{\cal T}_{23}(v)=
{\cal T}_{23}(v)
{\cal T}_{13}(u)R_{12}(u-v)\,,
\label{YBa}
\eeq
where the two auxiliary spaces are
${\cal V}_{1}=
{\cal V}_{2}=
{\bf C}^{k}$ and the quantum space is
${\cal V}_{3}= \otimes _{i=1}^{N}V_{i}$.

Graphically the ${\cal T}$-matrix looks as follows:

\vspace{5mm}

\special{em:linewidth 0.4pt}
\unitlength 1.00mm
\linethickness{0.4pt}
\begin{center}
\begin{picture}(94.33,14.00)
\emline{18.33}{6.33}{1}{58.67}{6.33}{2}
\put(60.67,6.33){\circle*{0.50}}
\put(63.67,6.33){\circle*{0.50}}
\put(67.00,6.33){\circle*{0.50}}
\put(94.33,6.33){\vector(1,0){0.2}}
\emline{70.00}{6.33}{3}{94.33}{6.33}{4}
\put(23.67,14.00){\vector(0,1){0.2}}
\emline{23.67}{4.00}{5}{23.67}{14.00}{6}
\put(33.67,14.00){\vector(0,1){0.2}}
\emline{33.67}{4.00}{7}{33.67}{14.00}{8}
\put(43.33,14.00){\vector(0,1){0.2}}
\emline{43.33}{4.00}{9}{43.33}{14.00}{10}
\put(53.67,14.00){\vector(0,1){0.2}}
\emline{53.67}{4.00}{11}{53.67}{14.00}{12}
\put(73.67,14.00){\vector(0,1){0.2}}
\emline{73.67}{4.00}{13}{73.67}{14.00}{14}
\put(83.67,14.00){\vector(0,1){0.2}}
\emline{83.67}{4.00}{15}{83.67}{14.00}{16}
\qbezier(23.67,9.33)(26.67,9.33)(26.67,6.33)
\qbezier(33.67,9.33)(36.67,9.33)(36.67,6.33)
\qbezier(43.67,9.33)(46.67,9.33)(46.67,6.33)
\qbezier(53.67,9.33)(56.67,9.33)(56.67,6.33)
\qbezier(73.67,9.33)(76.67,9.33)(76.67,6.33)
\qbezier(83.67,9.33)(86.67,9.33)(86.67,6.33)
\put(23.67,0.00){\makebox(0,0)[cc]{$y_1$}}
\put(33.67,0.00){\makebox(0,0)[cc]{$y_2$}}
\put(83.67,0.00){\makebox(0,0)[cc]{$y_N$}}
\put(13.33,6.33){\makebox(0,0)[cc]{$=$}}
\put(0.00,6.33){\makebox(0,0)[cc]{\Large{${\mathcal T}(u)$}}}
\end{picture}
\end{center}

\vspace{5mm}

\noindent
Traditionally, the auxiliary space is attached to the horizontal
line while the quantum space is assigned to the vertical lines
with rapidities $y_i$.

\subsubsection*{Fusion procedure}

Using the $R$-matrix (\ref{A1}) as a building block, it is possible
to construct more complicated solutions to the Yang-Baxter
equation. This is done by multiplying the fundamental solutions
and subsequent projection on irreducible representations of $GL(k)$.
We give the general scheme of the fusion in the auxiliary space
which is of prime importance for our purposes.

Let $\lambda =(\lambda _{1}, \lambda _{2}, \ldots , \lambda _{m})$
be a Young diagram with $n$ boxes and $m$ lines
$\lambda _{1}\geq \lambda _{2}\geq \ldots \geq \lambda _{m}$,
$m\leq k$, $\sum _{i=1}^{m}\lambda _{i}=n$.
Let
$$
P_{\lambda }:\otimes _{i=1}^{n}V_{i}\rightarrow V^{(\lambda )}
$$
be the projection operator on the space of the irreducible
representation of $GL(k)$ corresponding to $\lambda$. Write in the
box with coordinates $(i,j)$ ($i$-th line and $j$-th column)
the number
\beq
s_{(ij)}=u-2(i-j)\,.
\label{sij}
\eeq
For example:
\beq
\young{u&u+2&u+4&u+6\cr u-2&u&u+2\cr u-4\cr}
\label{ex1}
\eeq
Enumerating all the boxes in the natural order
(from left to right in the first line, then continue from left to
right in the second line and so on), one gets the ordered sequence
of numbers $s_1 , s_2 , \ldots , s_n$ (in the example above
$s_1 =u$, $s_2 =u+2$, $s_3 =u+4$, $s_4 =u+6$, $s_5 =u-2$,
$s_6 =u$, $s_7 =u+2$, $s_8 =u-4$).

The $R$-matrix acting in $V^{(\lambda )}\otimes {\bf C}^{k}$ is
\beq
R^{(\lambda )}(u)=P_{\lambda}R_{n0}(s_n )\otimes \ldots
\otimes R_{20}(s_2 ) \otimes
R_{10}(s_1 )P_{\lambda} \,.
\label{A4}
\eeq
The fundamental $R$-matrices $R_{i0}(s_i )$ are tensor multiplied
in the auxiliary spaces $V_i ={\bf C}^{k}$, their entries being
multiplied in the common quantum space
$V_0 ={\bf C}^{k}$ in the indicated order. Graphically,

\vspace{5mm}

\special{em:linewidth 0.4pt}
\unitlength 1.00mm
\linethickness{0.4pt}
\begin{center}
\begin{picture}(80.33,45.67)
\emline{25.33}{0.00}{1}{25.33}{40.00}{2}
\emline{25.33}{40.00}{3}{35.33}{40.00}{4}
\emline{35.33}{40.00}{5}{35.33}{0.00}{6}
\emline{35.33}{0.00}{7}{25.33}{0.00}{8}
\emline{35.33}{5.00}{9}{65.33}{5.00}{10}
\emline{35.33}{15.00}{11}{65.33}{15.00}{12}
\emline{35.33}{25.00}{13}{65.33}{25.00}{14}
\emline{35.33}{35.00}{15}{65.00}{35.00}{16}
\emline{65.33}{0.00}{17}{65.33}{40.00}{18}
\emline{65.33}{40.00}{19}{75.33}{40.00}{20}
\emline{75.33}{40.00}{21}{75.33}{0.00}{22}
\emline{75.33}{0.00}{23}{65.33}{0.00}{24}
\emline{20.33}{19.75}{27}{25.33}{19.75}{28}
\emline{20.33}{20.00}{27}{25.33}{20.00}{28}
\emline{20.33}{20.25}{27}{25.33}{20.25}{28}
\emline{75.33}{19.75}{29}{80.33}{19.75}{30}
\emline{75.33}{20.00}{29}{80.33}{20.00}{30}
\emline{75.33}{20.25}{29}{80.33}{20.25}{30}
\put(50.33,45.67){\vector(0,1){0.2}}
\emline{50.33}{0.00}{31}{50.33}{45.67}{32}
\put(50.33,5.00){\oval(3.00,3.00)[tr]}
\put(0.00,20.00){\makebox(0,0)[cc]{\Large{$R^{(\lambda)}(u)$}}}
\put(10.33,20.00){\makebox(0,0)[cc]{$=$}}
\put(30.00,20.00){\makebox(0,0)[cc]{\Large{$P_\lambda$}}}
\put(70.33,20.00){\makebox(0,0)[cc]{\Large{$P_\lambda$}}}
\put(60.33,7.67){\makebox(0,0)[cc]{$s_1$}}
\put(60.33,18.00){\makebox(0,0)[cc]{$s_2$}}
\put(60.33,38.67){\makebox(0,0)[cc]{$s_n$}}
\put(50.33,-5.00){\makebox(0,0)[cc]{$0$}}
\put(50.33,15.00){\oval(3.00,3.00)[tr]}
\put(50.33,25.00){\oval(3.00,3.00)[tr]}
\put(50.33,35.00){\oval(3.00,3.00)[tr]}
\end{picture}
\end{center}

\vspace{10mm}

The important property of $R^{(\lambda )}(u)$ is given by

\begin{th}
$R^{(\lambda )}(u)$ satisfies the Yang-Baxter equation in
${\cal V}_{1} \otimes
{\cal V}_{2} \otimes
{\cal V}_{3}$:
\beq
R_{12}^{(\lambda )}(u_1 -u_2 )
R_{13}^{(\lambda )}(u_1 -u_3 )
R_{23}(u_2 -u_3 )=
R_{23}(u_2 -u_3 )
R_{13}^{(\lambda )}(u_1 -u_3 )
R_{12}^{(\lambda )}(u_1 -u_2 )\,,
\label{YB1}
\eeq
where ${\cal V}_{1}=V^{(\lambda )}$,
${\cal V}_{2}={\cal V}_{3}={\bf C}^{k}$.
\end{th}

The idea of the proof is simple: to represent the projector
$P_{\lambda}$ as a product of fundamental $R$-matrices. This
representation alows one to move the lines using the Yang-Baxter
equation (\ref{YB}) for fundamental $R$-matrices. However,
implementation of this idea in general case is quite involved
from technical point of view. We find it useful to outline main steps
of the proof since it simultaneously helps to recover the structure
of zeros of the fused $R$-matrix, which will be important
later.

The following two elementary properties of the fundamental $R$-matrix
(\ref{A1}) are crucial for the construction:

a) There exists a value of $u$ at which $R(u)$ is proportional
to the permutation operator:
\beq
R_{12}(0)=2{\cal P}_{12}\,.
\label{A5}
\eeq
Graphicaly this means that if spectral parameters atached to
a pair of crossed lines coincide, one can "eliminate the intersection
point" as follows:

\vspace{5mm}

\special{em:linewidth 0.4pt}
\unitlength 1mm
\linethickness{0.4pt}
\begin{center}
\begin{picture}(72.33,40.00)
\emline{32.33}{20.00}{1}{50.33}{20.00}{2}
\emline{52.33}{22.00}{3}{52.33}{40.00}{4}
\emline{54.33}{19.67}{5}{70.33}{19.67}{6}
\emline{52.33}{17.67}{7}{52.33}{0.00}{8}
\put(50.33,22.00){\oval(4,4)[br]}
\put(54.33,17.67){\oval(4,4)[tl]}
\put(52.33,40.00){\vector(0,1){0.2}}
\emline{52.33}{40.00}{9}{52.33}{40.00}{10}
\put(70.33,19.67){\vector(1,0){0.2}}
\emline{72.00}{19.00}{11}{72.00}{19.00}{12}
\emline{53.33}{19.00}{13}{53.33}{19.00}{14}
\emline{52.33}{20.00}{15}{52.33}{20.00}{16}
\put(34.67,24.00){\makebox(0,0)[cc]{\large{$u$}}}
\put(56.33,0.33){\makebox(0,0)[cc]{\large{$u$}}}
\put(27.33,19.67){\makebox(0,0)[cc]{\large{$2$}}}
\put(20.00,19.67){\makebox(0,0)[cc]{$=$}}
\put(5.00,19.33){\makebox(0,0)[cc]{\Large{$R_{12}(0)$}}}
\end{picture}
\end{center}

\vspace{5mm}

b) The $R$-matrix degenerates at the points $\pm 2$:
\beq
R_{12}(\pm 2)=4P_{12}^{\pm}\,,
\label{A6}
\eeq
where $P_{12}^{\pm}$ progects on symmetric (antisymmetric) tensors
in $V_1 \otimes V_2$.

Recall the ordered sequence of spectral parameters $s_1 , s_2 ,
\ldots , s_n$ introduced at the beginning of this section. The
property b) suggests to try
\beq
P_{\lambda}^{naive}\propto \prod _{i<j}^{n}R_{ij}(s_i -s_j )\,,
\label{A7}
\eeq
where the structure of the ordered product corresponds to the pattern
of complete intersection of $n$ lines carrying spectral parameters
$s_i$. The order of successive intersections is irrelevant by
virtue of the Yang-Baxter equation. For example ($n=3$):

\vspace{5mm}

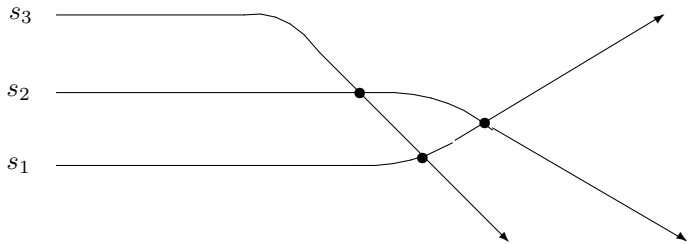
\special{em:linewidth 0.4pt}
\unitlength 1.00mm
\linethickness{0.4pt}
\begin{center}
\begin{picture}(88.67,31.00)
\put(88.67,0.00){\vector(3,-2){0.2}}
\emline{63.00}{15.00}{3}{88.67}{0.00}{4}
\put(85.67,30.00){\vector(3,2){0.2}}
\emline{58.00}{13.33}{5}{85.67}{30.00}{6}
\emline{30.00}{30.00}{7}{32.13}{30.13}{8}
\emline{32.13}{30.13}{9}{34.16}{29.73}{10}
\emline{34.16}{29.73}{11}{36.10}{28.82}{12}
\emline{36.10}{28.82}{13}{37.93}{27.40}{14}
\emline{37.93}{27.40}{15}{40.00}{25.00}{16}
\emline{30.00}{30.00}{17}{5.00}{30.00}{18}
\put(65.00,0.00){\vector(1,-1){0.2}}
\emline{40.00}{25.00}{19}{65.00}{0.00}{20}
\emline{50.00}{19.67}{21}{52.49}{19.44}{22}
\emline{52.49}{19.44}{23}{54.85}{18.96}{24}
\emline{54.85}{18.96}{25}{57.09}{18.25}{26}
\emline{57.09}{18.25}{27}{59.19}{17.30}{28}
\emline{59.19}{17.30}{29}{61.16}{16.10}{30}
\emline{61.16}{16.10}{31}{63.00}{14.67}{32}
\emline{5.00}{19.67}{33}{50.00}{19.67}{34}
\emline{45.00}{10.00}{35}{47.46}{10.00}{36}
\emline{47.46}{10.00}{37}{49.83}{10.23}{38}
\emline{49.83}{10.23}{39}{52.12}{10.70}{40}
\emline{52.12}{10.70}{41}{54.32}{11.40}{42}
\emline{54.32}{11.40}{43}{57.67}{13.00}{44}
\emline{5.00}{10.00}{45}{45.00}{10.00}{46}
\put(45.33,19.67){\circle*{1.50}}
\put(53.67,11.00){\circle*{1.50}}
\put(62.00,15.67){\circle*{1.50}}
\put(0.33,30.00){\makebox(0,0)[cc]{$s_3$}}
\put(0.00,19.67){\makebox(0,0)[cc]{$s_2$}}
\put(0.00,10.00){\makebox(0,0)[cc]{$s_1$}}
\end{picture}
\end{center}

\vspace{5mm}

\noindent
One can prove by induction that this prescription does work for
"hook diagrams" $\lambda = (\lambda _{1}, 1, 1, \ldots , 1)$
(i.e. such that any diagonal contains exactly one box) providing
the desired projection operator.

However, in general case we encounter a difficulty:
$P_{\lambda }^{naive}$ can be identically zero. Indeed, if
$\lambda = (2,2)$ we have

\vspace{5mm}

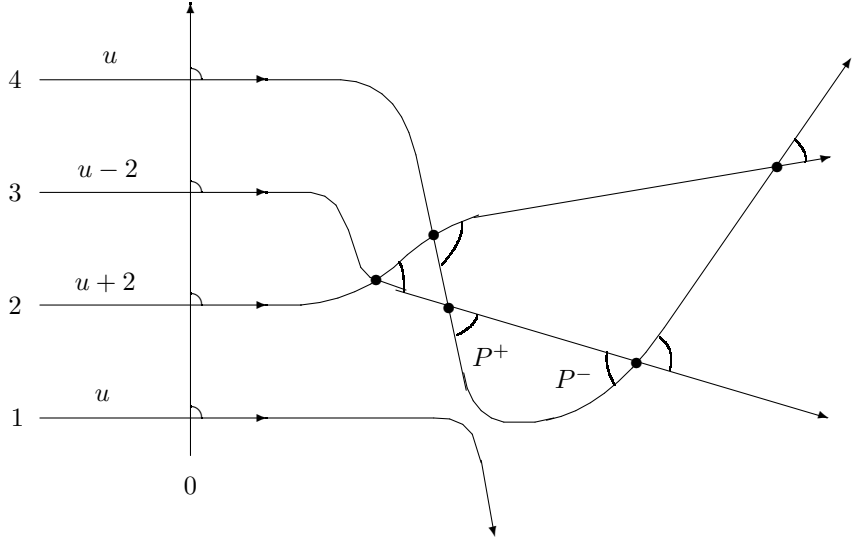
\special{em:linewidth 0.4pt}
\unitlength 1mm
\linethickness{0.4pt}
\begin{center}
\begin{picture}(112.67,75.00)
\put(35.00,65.00){\vector(1,0){0.2}}
\emline{5.00}{65.00}{1}{35.00}{65.00}{2}
\emline{35.00}{65.00}{3}{45.00}{65.00}{4}
\emline{45.00}{65.00}{5}{47.03}{64.73}{6}
\emline{47.03}{64.73}{7}{48.85}{64.09}{8}
\emline{48.85}{64.09}{9}{50.46}{63.10}{10}
\emline{50.46}{63.10}{11}{51.87}{61.75}{12}
\emline{51.87}{61.75}{13}{53.08}{60.04}{14}
\emline{53.08}{60.04}{15}{54.07}{57.96}{16}
\emline{54.07}{57.96}{17}{55.00}{55.00}{18}
\put(25,65){\oval(3,3)[tr]}
\put(25,50){\oval(3,3)[tr]}
\put(25,35){\oval(3,3)[tr]}
\put(25,20){\oval(3,3)[tr]}
\put(35.00,50.00){\vector(1,0){0.2}}
\emline{5.00}{50.00}{19}{35.00}{50.00}{20}
\emline{35.00}{50.00}{21}{41.00}{50.00}{22}
\emline{41.00}{50.00}{23}{42.86}{49.48}{24}
\emline{42.86}{49.48}{25}{44.35}{48.29}{26}
\emline{44.35}{48.29}{27}{46.00}{45.00}{28}
\emline{46.00}{45.00}{29}{47.67}{40.00}{30}
\emline{47.67}{40.00}{31}{48.89}{38.70}{32}
\emline{48.89}{38.70}{33}{53.67}{37.00}{34}
\put(109.67,20.00){\vector(3,-1){0.2}}
\emline{52.33}{37.00}{35}{109.67}{20.00}{36}
\put(35.00,35.00){\vector(1,0){0.2}}
\emline{5.00}{35.00}{37}{35.00}{35.00}{38}
\emline{35.00}{35.00}{39}{39.67}{35.00}{40}
\emline{39.67}{35.00}{41}{39.67}{35.00}{42}
\emline{39.67}{35.00}{43}{42.21}{35.36}{44}
\emline{42.21}{35.36}{45}{44.61}{35.94}{46}
\emline{44.61}{35.94}{47}{46.85}{36.77}{48}
\emline{46.85}{36.77}{49}{48.95}{37.82}{50}
\emline{48.95}{37.82}{51}{52.00}{40.00}{52}
\emline{52.00}{40.00}{53}{53.83}{41.69}{54}
\emline{53.83}{41.69}{55}{55.74}{43.19}{56}
\emline{55.74}{43.19}{57}{57.74}{44.50}{58}
\emline{57.74}{44.50}{59}{59.82}{45.61}{60}
\emline{59.82}{45.61}{61}{63.33}{47.00}{62}
\put(110.00,54.67){\vector(4,1){0.2}}
\emline{62.67}{46.67}{63}{110.00}{54.67}{64}
\put(35.00,20.00){\vector(1,0){0.2}}
\emline{5.00}{20.00}{65}{35.00}{20.00}{66}
\put(25.00,75.00){\vector(0,1){0.2}}
\emline{25.00}{15.00}{67}{25.00}{75.00}{68}
\put(57.33,44.33){\circle*{1.50}}
\put(49.67,38.33){\circle*{1.50}}
\put(59.33,34.67){\circle*{1.50}}
\put(84.33,27.33){\circle*{1.50}}
\emline{55.00}{55.00}{69}{61.67}{23.67}{70}
\emline{57.33}{20.00}{71}{59.48}{19.85}{72}
\emline{59.48}{19.85}{73}{61.19}{19.12}{74}
\emline{61.19}{19.12}{75}{62.46}{17.81}{76}
\emline{62.46}{17.81}{77}{63.67}{14.00}{78}
\emline{35.00}{20.00}{79}{57.33}{20.00}{80}
\emline{61.33}{26.00}{81}{61.81}{23.94}{82}
\emline{61.81}{23.94}{83}{62.58}{22.25}{84}
\emline{62.58}{22.25}{85}{63.65}{20.94}{86}
\emline{63.65}{20.94}{87}{65.00}{20.00}{88}
\emline{65.00}{20.00}{89}{66.65}{19.44}{90}
\emline{66.65}{19.44}{91}{70.81}{19.44}{92}
\emline{70.81}{19.44}{93}{73.33}{20.00}{94}
\emline{72.33}{19.67}{95}{74.30}{20.19}{96}
\emline{74.30}{20.19}{97}{76.24}{20.97}{98}
\emline{76.24}{20.97}{99}{78.16}{22.01}{100}
\emline{78.16}{22.01}{101}{80.04}{23.29}{102}
\emline{80.04}{23.29}{103}{81.90}{24.84}{104}
\emline{81.90}{24.84}{105}{83.73}{26.64}{106}
\emline{83.73}{26.64}{107}{85.53}{28.69}{108}
\emline{85.53}{28.69}{109}{88.00}{32.00}{110}
\put(112.67,67.67){\vector(2,3){0.2}}
\emline{88.00}{32.00}{111}{112.67}{67.67}{112}
\put(103.00,53.33){\circle*{1.50}}
\put(65.33,5.00){\vector(1,-4){0.2}}
\emline{63.67}{14.33}{113}{65.33}{5.00}{114}
\put(2.00,20.00){\makebox(0,0)[cc]{1}}
\put(1.67,35.00){\makebox(0,0)[cc]{2}}
\put(1.67,50.00){\makebox(0,0)[cc]{3}}
\put(1.67,65.00){\makebox(0,0)[cc]{4}}
\put(14.33,68.00){\makebox(0,0)[cc]{$u$}}
\put(14.00,53.00){\makebox(0,0)[cc]{$u-2$}}
\put(13.67,38.00){\makebox(0,0)[cc]{$u+2$}}
\put(13.33,23.00){\makebox(0,0)[cc]{$u$}}
\put(65.00,28.33){\makebox(0,0)[cc]{$P^+$}}
\put(76.00,25.33){\makebox(0,0)[cc]{$P^-$}}
\qbezier(52.67,40.67)(53.33,40.00)(53.33,36.67)
\qbezier(61.00,46.00)(61.33,43.67)(58.33,40.33)
\qbezier(63.00,33.67)(63.00,32.33)(60.33,31.00)
\qbezier(80.33,28.67)(80.00,26.67)(81.33,24.33)
\qbezier(87.33,30.67)(89.33,29.33)(88.67,26.33)
\qbezier(106.67,54.00)(107.00,55.33)(105.33,57.00)
\put(25.00,11.00){\makebox(0,0)[cc]{0}}
\end{picture}
\end{center}

\vspace{5mm}

\noindent
that is zero since the
complementary projectors $P^+$ and $P^-$ meet
together. A little inspection shows that this zero arises
for the same reason each time when there are at least two lines
$i\neq j$ such that $s_i =s_j$. Such lines always exist if the
Young diagram contains a diagonal with at least two boxes. In this
case a sort of regularization is necessary.

Let us introduce a small parameter $\epsilon$ and modify the
sequence $s_1, s_2 , \ldots , s_n$ writing
\beq
s_{(ij)}^{\epsilon}=u-2(i-j)+(i-1)\epsilon
\label{A8}
\eeq
in the box $(i,j)$ with the same linear order $s_{1}^{\epsilon},
s_{2}^{\epsilon}, \ldots , s_{n}^{\epsilon}$. Consider the operator
\beq
P_{\lambda}^{(\epsilon)}=\prod _{i,j}^{n}R_{ij}
(s_{i}^{\epsilon}-s_{j}^{\epsilon})
\label{A9}
\eeq
with the same order in the product as in eq.\,(\ref{A7}).

\begin{lem}
As $\epsilon \rightarrow 0$,
\beq
P_{\lambda }^{(\epsilon)}=c\epsilon ^{\kappa}P_{\lambda}+
{\cal O}(\epsilon ^{\kappa +1})\,,
\label{A10}
\eeq
where $\kappa = \#\{(i_1 , j_1 ), (i_2 , j_2 )|i_1 <i_2 , \;
i_1 - j_1 = i_2 - j_2 \}$ is the number of pairs of boxes on the
diagonal lines and $c$ is a numerical constant.
\end{lem}

This lemma is sufficient to prove the Yang-Baxter equation
for $R^{(\lambda )}(u)$.

The fusion procedure in the quantum space is made in a similar way.
This procedure allows one to define the $R$-matrix
$R^{(\lambda)(\mu)}(u)$ acting in $V^{(\lambda )}\otimes
V^{(\mu )}$ for two arbitrary Young diagrams $\lambda$, $\mu$.
The Yang-Baxter equation then reads
\beq
R_{12}^{(\lambda )(\mu )}(u_1 -u_2 )
R_{13}^{(\lambda )}(u_1 -u_3 )
R_{23}^{(\mu )}(u_2 -u_3 )=
R_{23}^{(\mu )}(u_2 -u_3 )
R_{13}^{(\lambda )}(u_1 -u_3 )
R_{12}^{(\lambda )(\mu )}(u_1 -u_2 )\,,
\label{YB2}
\eeq
where ${\cal V}_{1}=V^{( \lambda )}$,
${\cal V}_{2}=V^{( \mu )}$,
${\cal V}_{3}={\bf C}^{k}$.

Each matrix element of $R^{(\lambda )}(u)$ is a polynomial in $u$.
However, the fusion procedure brings to
$R^{(\lambda )}(u)$ a number of "trivial" zeros in the sense that
they are common for all the matrix elements. In other words,
there are some values of $u$ at which
$R^{(\lambda )}(u)$ is the zero operator. These zeros should be
extracted.

Using the same argument as above, it is easy to see that
\beq
R^{(\lambda )}(u-s_i )=0, \;\;\;\;\;\;i=2,3,\ldots, n\,,
\label{A11}
\eeq
the remaining $n$-th zero being non-trivial. If $s_i =s_j$
for $i\neq j$, we have zero of higher degree. Extracting the common
multiplier, we redefine
\beq
R^{(\lambda )}(u )\rightarrow
R^{(\lambda )}(u )\left ( \prod _{i=2}^{n}s_i \right )^{-1}
\label{A12}
\eeq
that is a polynomial of degree $1$ with operator coefficients.

At last, we specialize this formula for rectangular Young diagrams
of length $s$ and height $a$: $\lambda =(s,s, \ldots , s)$ ($a$ times):
\beq
R^{(a\times s )}(u )\rightarrow
R^{(a\times s )}(u )\left (
\prod _{q=0}^{a-1}
\prod _{p=1}^{s-1}(u-2q+2p) \right )^{-1} \left (
\prod _{q=1}^{a-1} (u-2q)
\right )^{-1}.
\label{A13}
\eeq
For $a=1$ the last product should be skiped, for $s=1$ the double
product should be skipped.

\subsubsection*{Fused ${\cal T}$-matrices}

Similarly to eq.\,(\ref{A3}) one can introduce the
${\cal T}$-matrix
with the auxiliary space $V_{0}=V^{(\lambda )}$ and the quantum space
$\otimes _{i=1}^{N}V_{i}$:
\beq
{\cal T}^{(\lambda )}(u)=R_{0N}^{(\lambda )}(u-y_{N})\ldots
R_{02}^{(\lambda )}(u-y_{2})
R_{01}^{(\lambda )}(u-y_{1})\,.
\label{A14}
\eeq
It follows from (\ref{YB2}) and (\ref{A14}) that
\beq
R_{12}^{(\lambda )(\mu )}(u-v){\cal T}_{13}^{(\lambda )}(u)
{\cal T}_{23}^{(\mu )}(v)=
{\cal T}_{23}^{(\mu )}(v)
{\cal T}_{13}^{(\lambda )}(u)R_{12}^{(\lambda )(\mu )}(u-v)\,,
\label{YB3}
\eeq
where the two auxiliary spaces are
${\cal V}_{1}= V^{(\lambda )}$,
${\cal V}_{2}= V^{(\mu )}$,
and the quantum space is
${\cal V}_{3}= \otimes _{i=1}^{N}V_{i}$.

Each factor in (\ref{A14}) gives an independent contribution to the
common polynomial multiplier. Let us explicitly extract this
multiplier in the case of rectangular Young diagrams $a\times s$:
\beq
{\cal T}^{(a\times s )}(u )=
\mbox{{\bf T}}^{(a\times s )}(u )
\left (
\prod _{q=0}^{a-1}
\prod _{p=1}^{s-1}\phi (u-2q+2p) \right ) \left (
\prod _{q=1}^{a-1} \phi (u-2q)
\right ).
\label{A15}
\eeq
Here
\beq
\phi (u)=\prod _{j=1}^{N}(u-y_{j})
\label{A16}
\eeq
and
$\mbox{{\bf T}}^{(a\times s )}(u)$ denotes the
renormalized ${\cal T}$-matrix. For generic polynomial
$\phi (u)$ it
contains non-trivial "operator" zeros only.
If there are pairs $i,j$ such that $y_i -y_j =2$,
$\mbox{{\bf T}}^{(a\times s )}(u)$ acquires some extra "trivial"
zeros and should be renormalized further. This case corresponds to
higher representations in the quantum space.

\subsubsection*{Quantum transfer matrices}

The key notion of the theory is
{\it quantum transfer matrix}\footnote{The name "transfer matrix"
came from statistical mechanics on the lattice. It is a bit misleading
since the transfer matrix is a {\it scalar} in the auxiliary space. The
word "matrix" here is related to the quantum space.} obtained by
taking trace of ${\cal T}^{(\lambda )}(u)$ in the auxiliary space:
\beq
T^{(\lambda )}(u)=\mbox{Tr}_{aux}
{\cal T}^{(\lambda )}(u-\lambda _{1}
+\lambda _{1}')\,,
\label{A17}
\eeq
where $\lambda '$ denotes the transposed diagram (i.e. reflected
with respect to the main diagonal), so $\lambda _{1}'$ is height
of the first column of $\lambda $. The shift of the spectral parameter
is introduced for later convenience.

The crucial property of quantum transfer matrices is their
commutativity for all values of the continuous spectral parameter $u$
and the discrete spectral parameter $\lambda $:
\beq
\phantom{a}[T^{(\lambda )}(u),\;
T^{(\mu )}(v)]=0\,.
\label{A18}
\eeq
This fact immediately follows from the Yang-Baxter relation (\ref{YB3}).

Therefore, we have a family of commuting operators in the quantum
space. The point is that hamiltonians of quantum integrable systems
belong to this family, so $T^{( \lambda )}(u)$ are integrals of motion.
In other words,
$T^{( \lambda )}(u)$ is a generating function for commuting
hamiltonians.
Simultaneous diagonalization of transfer
matrices
$T^{(\lambda )}(u)$ is one of the corner-stones of the theory. In this
way one obtains spectral characteristics of quantum integrable systems.

For the case of rectangular diagrams $a\times s$ we introduce the
special notation
\beq
T_{s}^{a}(u)\equiv T^{(a\times s)}(u)
\label{A19}
\eeq
which will be used in the sequel.

Extracting the "trivial" zeros (\ref{A15}) amounts to changing the
normalization:
\beq
\label{A20}
T_{s}^{a}(u)\rightarrow T_{s}^{a}(u)
\left (\prod _{l=0}^{a-1}\prod _{p=1}^{s-1}
\phi (u-s-a+2l+2p+2)
\prod _{l=1}^{a-1}\phi (u-s-a+2l)\right )^{-1}\,.
\eeq
Each eigenvalue of the transfer matrix is a polynomial in $u$
and all the "trivial" zeros
(common for all the eigenvalues) are removed.

\subsubsection*{Quantum determinant}

The maximal "external power" of the fundamental ${\cal T}$-matrix
${\cal T}(u)$ in the auxiliary space ${\bf C}^{k}$ is called the
{\it quantum determinant} of ${\cal T}(u)$. More precisely, we put
\beq
D^{(k)}(u)\equiv \det _{q}{\cal T}(u) \equiv T_{1}^{k}(u)=
{\cal T}^{(1^{k})}(u-1+k)\,.
\label{qdet1}
\eeq
The last equality is due to the fact that the fused auxiliary space is
one-dimensional. In other words, $D^{(k)}(u)$ is a scalar in the auxiliary
space by definition. The following theorem shows that
$D^{(k)}(u)$ is a scalar in the quantum space too.

\begin{th}
The quantum determinant
$D^{(k)}(u)$ of the fundamental
${\cal T}$-matrix
${\cal T}(u)$ with $k$-dimensional auxiliary space lies in the center
of the algebra generated by matrix elements of
${\cal T}(u)$:
\beq
\phantom{a}[D^{(k)}(u),\;({\cal T}(v))_{ij}]=0\,, \;\;\;\;\;\;
i,j=1,2, \ldots , k\,.
\label{qdet2}
\eeq
It is given explicitly by the formula
\beq
D^{(k)}(u)=\phi (u+k+1)\prod _{l=1}^{k-1}\phi (u-k-1+2l)\,,
\label{qdet3}
\eeq
where $\phi (u)$ is defined in eq.\,(\ref{A16}).
\end{th}
The idea of the proof is to use the Yang-Baxter equation which
allows one to reduce everything to the quantum determinant of the
fundamental $R$-matrix. In this case the assertion can be verified
by a direct computation.

Let us show how to verify eq.\,(\ref{qdet3}) without any computation.
The last $k-1$ factors contain just the "trivial" zeros coming from
eq.\,(\ref{A20}), so we only need to explain the first factor. Note
that at $u=2$ the fused $R$-matrix contains the following block:

\vspace{5mm}

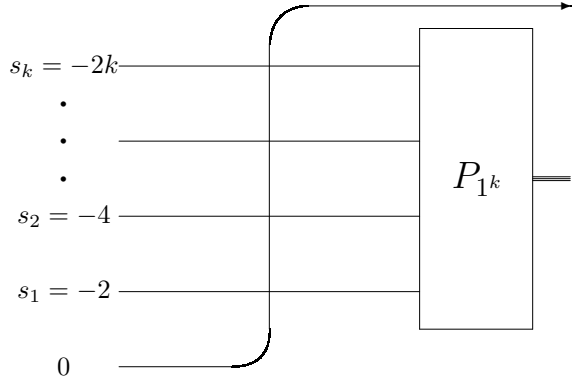
\special{em:linewidth 0.4pt}
\unitlength 1mm
\linethickness{0.4pt}
\begin{center}
\begin{picture}(75.00,53.00)
\emline{15.00}{45.00}{1}{55.00}{45.00}{2}
\emline{15.00}{35.00}{3}{55.00}{35.00}{4}
\emline{15.00}{25.00}{5}{55.00}{25.00}{6}
\emline{15.00}{15.00}{7}{55.00}{15.00}{8}
\emline{55.00}{50.00}{9}{55.00}{10.00}{10}
\emline{55.00}{10.00}{11}{70.00}{10.00}{12}
\emline{70.00}{10.00}{13}{70.00}{50.00}{14}
\emline{70.00}{50.00}{15}{55.00}{50.00}{16}
\emline{70.00}{29.75}{17}{75.00}{29.75}{18}
\emline{70.00}{30.00}{17}{75.00}{30.00}{18}
\emline{70.00}{30.25}{17}{75.00}{30.25}{18}
\emline{35.00}{15.00}{19}{35.00}{48.00}{20}
\put(75.00,53.00){\vector(1,0){0.2}}
\emline{40.00}{53.00}{21}{75.00}{53.00}{22}
\qbezier(35.00,48.00)(35.33,52.67)(40.00,53.00)
\emline{35.00}{15.00}{29}{35.00}{10.00}{30}
\emline{30.00}{5.00}{31}{15.00}{5.00}{32}
\qbezier(30.00,5.00)(35.00,5.00)(35.00,10.00)
\put(7.67,5.00){\makebox(0,0)[cc]{0}}
\put(7.67,15.00){\makebox(0,0)[cc]{$s_1=-2$}}
\put(7.67,25.00){\makebox(0,0)[cc]{$s_2=-4$}}
\put(7.67,45.00){\makebox(0,0)[cc]{$s_k=-2k$}}
\put(62.67,30.00){\makebox(0,0)[cc]{\Large{$P_{1^k}$}}}
\put(7.67,30.00){\circle*{0.67}}
\put(7.67,35.00){\circle*{0.67}}
\put(7.67,40.00){\circle*{0.67}}
\end{picture}
\end{center}

\vspace{5mm}

\noindent
which is zero since it is the projector on the $(k+1)$-th external
power of ${\bf C}^{k}$. So $R^{(1^{k})}(2)=0$ as the whole thing.
This yields the first factor in eq.\,(\ref{qdet3}).

There is a more general formula derived by the same argument,
\beq
\label{qdet4}
T_{s}^{k}(u)=\phi (u+s+k)\left (\prod_
{l=0}^{k-1}\prod _{p=1}^{s-1} \phi (u-s-k+2l+2p+2) \right )
\prod _{l=1}^{k-1}\phi
(u-s-k+2l)\,,
\;\;\;\;T_{s}^{0}(u)=1\,.
\eeq
which will be used in Sect.\,4.

\subsubsection*{Functional relations}

A combination of the fusion procedure and the Yang-Baxter equation results
in numerous functional relations (fusion rules) for quantum
transfer matrices.

We illustrate the origin of these relations on the simplest example.
Consider the product
\beq
T_{1}^{1}(u-2)
T_{1}^{1}(u)=\mbox{Tr}_{V_{1}\otimes V_{2}}
\big ( {\cal T}_{20}(u-2)
\otimes
{\cal T}_{10}(u)\big )\,,
\label{A21}
\eeq
where $V_{1}=V_{2}={\bf C}^{k}$ are auxiliary spaces and the quantum
space $V_{0}$ is arbitrary. Insert the unit matrix $I=P^{+}+P^{-}$
represented as sum of the complementary projectors inside the trace.
Using cyclicity of the trace, the property $(P^{\pm })^2 =P^{\pm }$ and
the Yang-Baxter equation, we arrive at the relation
\beq
T_{1}^{1}(u-2)
T_{1}^{1}(u)=T_{1}^{2}(u-1) + T_{2}^{1}(u-1)\,.
\label{A22}
\eeq
A more general relation derived in a similar way is
\beq
T_{s}^{1}(u+1)
T_{1}^{1}(u-s)=T^{(\lambda =(s,1))}(u) + T_{s+1}^{1}(u)\,.
\label{A23}
\eeq

Proceeding further, it is possible to show that $T^{(\lambda )}(u)$
can be expressed through either $T_{s}^{1}(u)$ or $T_{1}^{a}(u)$ only.

\begin{th}
The following determinant formulas hold true:
\beq
T^{(\lambda )}(u)=
\det _{1\leq i,j\leq \lambda _{1}'}\big ( T_{\lambda _{i}-i+j}^{1}
(u-\lambda _{1}'+\lambda _{1}-\lambda _{i}+i+j-1)\big )\,,
\label{det1}
\eeq
\beq
T^{(\lambda )}(u)=
\det _{1\leq i,j\leq \lambda _{1}}\big ( T_{1}^{\lambda _{i}'-i+j}
(u-\lambda _{1}'+\lambda _{1}+\lambda _{i}'-i-j+1)\big )\,,
\label{det2}
\eeq
where it is implied that $T^{(\emptyset )}(u)=1$ ($\emptyset $ denotes the
empty diagram).
\end{th}
We remind the reader that $\lambda '$ denotes the transposed diagram. The
entries $T_{m}^{1}(u)$ (resp., $T_{1}^{m}(u)$) of the matrix in
eq.\, (\ref{det1}) (resp., (\ref{det2})) are transfer matrices corresponding
to the one-line (resp., one-column) diagrams.

These formulas were obtained by V.Bazhanov and N.Reshetikhin.
Sometimes they are called quantum Jacobi-Trudi formulas. They are
"Yang-Baxterization" of the classical Jacobi-Trudi identities in the
sense that the latter are reproduced from the former by forgetting
the dependence on $u$.

The specification of eqs.\,(\ref{det1}), (\ref{det2}) to rectangular
diagrams reads
\beq
T_{s}^{a}(u)=\det \big (T^{1}_{s+i-j}(u-i-j+a+1)\big ),
\;\;\;\;\; i,j=1,\ldots ,a\,, \;\;\;\;\; T_{s}^{0}(u)=1\,,
\label{det3}
\eeq
\beq
T_{s}^{a}(u)=\det \big (T_{1}^{a+i-j}(u-i-j+s+1)\big ),
\;\;\;\;\; i,j=1,\ldots ,s\,, \;\;\;\;\; T^{a}_{0}(u)=1\,.
\label{det4}
\eeq

These determinant formulas exhibit a very nice structure.
Though, they give only a partial solution because
$T_{s}^{1}(u)$ or
$T_{1}^{a}(u)$ (entering as "initial data") are still to be
determined.

\subsubsection*{Bilinear form of the fusion rules}

The fusion rules can be recast into another suggestive form  which is
in a sense more practically useful since it allows one to obtain
a complete solution. At the same time, this form provides a link with
classical non-linear integrable equations, which is our main concern here.

It follows from eqs.\,(\ref{det3}), (\ref{det4}) that transfer
matrices for {\it rectangular} Young diagrams obey a closed set of
relations among themselves. Using the Jacobi identity for
determinants, they can be represented in the following
model-independent bilinear form:
\beq
T^{a}_{s}(u+1)T^{a}_{s}(u-1)-
T^{a}_{s+1}(u)T^{a}_{s-1}(u)=
T^{a+1}_{s}(u)T^{a-1}_{s}(u)\,.
\label{A24}
\eeq
Since $T^{a}_{s}(u)$ commute at different $u,\,a,\,s,$, the same equation
holds for all eigenvalues of the
transfer matrix, so from now on we can (and will) treat
$T^{a}_{s}(u)$ as a number-valued function. Note that the 3 variables
$u,s,a$ enter almost symmetricallyin spite of their very
different nature.

Let us mention an analogue of (\ref{A24}) for more complicated diagrams.
Consider,
for example, Young diagrams consisting of two rectangular
blocks (i.e. with $a_1$ lines of length $s_1 +s_2$ and the rest
$a_2$ lines of length $s_1$) and let $T_{s_{1}, s_{2}}^{a_1 , a_2 }(u)$
be the corresponding transfer matrix. Then it holds
\begin{eqnarray}
&&T_{s_1 , s_2 }^{a_1 , a_2 -1}(u)
T_{s_1 -1, s_2 -1}^{a_1 , a_2 +1}(u)+
T_{s_1 , s_2 +1}^{a_1 -1, a_2 -1}(u)
T_{s_1 , s_2 -1}^{a_1 +1, a_2 +1}(u) \nonumber \\
&+&T_{s_1 +1, s_2 }^{a_1 -1, a_2 }(u+1)
T_{s_1 -1, s_2 }^{a_1 +1, a_2 }(u-1) \nonumber \\
&=&T_{s_1 +1, s_2 }^{a_1 , a_2 -1}(u+1)
T_{s_1 -1, s_2 }^{a_1 , a_2 +1}(u-1)+
T_{s_1 , s_2 +1}^{a_1 -1, a_2 }(u+1)
T_{s_1 , s_2 -1}^{a_1 +1, a_2 }(u-1)\,.
\label{A25}
\end{eqnarray}
In what follows we mainly deal with eq.\,(\ref{A24}) only.

Equation (\ref{A24}) is the top of the {\it quantum} construction
described in this section. Remarkably, this very equation is well known
in the theory of {\it classical} soliton equations. This is famous
Hirota's bilinear difference equation (HBDE) for the function
$T_{s}^{a}(u)$ of 3 variables. The reason of this coincidence is
hidden somewhere in the representation theory. We believe that there
should exist a formulation of the quantum theory such that this
coincidence is obvious from the very beginning. However,
we are not going to dwell upon this aspect of the problem.
Our aim here is more technical. In the sequel, we treat the
fundamental relation (\ref{A24}) not as an identity but as a
{\it fundamental
equation} and show how to extract from it the specific quantum
information. The solution to
(classical!) HBDE then appears in the form of Bethe
equations. In our opinion, this fact is quite remarkable by itself.
On the other hand, we anticipate that this approach makes it
possible to use some specific tools of classical integarbility
in quantum problems.

\subsection*{Comments and references}

1. Throughout this section we use notations and terminology of the
quantum inverse scattering method (QISM) developed by the former
Leningrad school. At present QISM ($=$ algebraic Bethe ansatz)
provides the most convenient framework to analyse the phenomenon
of quantum integrability. The early paper \cite{FadTakh} remains one
of the best reviews of the subject; see also the book \cite{book}.
The graphical interpretation is based on the reformulation in terms
of the factorized scattetring \cite{factorized}.

The fusion procedure was invented by P.Kulish, N.Reshetikhin and
E.Sklyanin \cite{KRS}, \cite{GL3} and generalized to elliptic solutions
of the Yang-Baxter equation by I.Cherednik \cite{Cher1}. The
regularization (\ref{A8}) was suggested in ref.\,\cite{Cher2}, for the
general proof of the analogue of Lemma 1.1 in the elliptic case see
\cite{JKMO}.

The notion of quantum determinant was introduced in the paper
\cite{IzKor} for $2\times 2$ monodromy matrices and generalized to
$k\times k$ matrices in the paper \cite{KRS}. For the complete
proof of Theorem 1.2 in the elliptic case see e.g.\,\cite{Haseg}.

2. Functional relations for transfer matrices have a long story.
Some of them appeared for the first time in the papers \cite{GL3},
\cite{Reshet}. The general quantum Jacobi-Trudi formulas (\ref{det1}),
(\ref{det2}) appeared in \cite{BR}, see also \cite{KOS}. The bilinear
form of functional relations was first suggested by A.Klumper and
P.Pearce in the particular case of $A_1$-type models and then
generalized to the $A_{k-1}$-case by A.Kuniba, T.Nakanishi and
J.Suzuki \cite{Kuniba1}. They also suggested bilinear fusion relations
for models associated to other Dynkin graphs. Eq.\,(\ref{A25})
and its generalizations to other types of Yang diagrams can be
obtained \cite{KLWZ} from determinant formulas (\ref{det1}),
(\ref{det2}) by means of the Pl\"ucker relations.

3. The transfer matrices play the role of quantum characters.
They may be also called "quantum Schur functions". Emphasizing this
analogy, we use notations and conventions from the book \cite{Macdonald}
on symmetric functions.

4. The results described in this section are valid for elliptic
solutions of the Yang-Baxter equation, too (sometimes with minor
changes). Instead of Yang's $R$-matrix (\ref{A2}) one should start
from Belavin's $R$-matrix \cite{Belavin} with elliptic dependence
on the spectral parameter:
\beq
R(u)=\sum R_{ij}^{i'j'}(u)E_{ii'}\otimes E_{jj'},\,
\label{ell1}
\eeq
where
\beq
R_{i'j'}^{ij}(u)=\frac{
\theta \left [
\begin{array}{c} \frac{j-i}{k}+\frac{1}{2}\\
\frac{1}{2}
\end{array}
\right ]
\big ( \eta (u+2)\big | k\tau \big )}
{
\theta \left [
\begin{array}{c} \frac{i'-i}{k}+\frac{1}{2}\\
\frac{1}{2}
\end{array}
\right ]
\big (2\eta \big | k\tau \big )
\theta \left [
\begin{array}{c} \frac{j-i'}{k}+\frac{1}{2}\\
\frac{1}{2}
\end{array}
\right ]
\big ( \eta u \big | k\tau \big )}
\label{ell2}
\eeq
if $i+j=i'+j' (\mbox{mod}\, k)$ and
$R_{i'j'}^{ij}(u)=0$ otherwise. Here
\beq
\theta \left [
\begin{array}{c} \alpha \\
\beta
\end{array}
\right ]
\big ( u\big | \tau \big )=
\sum _{l\in {\bf Z}}\exp \big (i\pi \tau (l+\alpha )^2
+2i\pi (l+\alpha )(u+\beta )\big )
\label{ell3}
\eeq
is the Jacobi theta-function with characteristics. There are two
parameters: $\eta$ and $\tau$. (This form of Belavin's $R$-matrix
is taken from refs.\,\cite{RT}, \cite{HZ}.)

Again, the $R$-matrix degenerates at $u=\pm 2$:
$R(-2)=M_{-}P^{-}$,
$R(+2)=P^{+}M_{+}$, where $M_{\pm }$ are some invertible matrices.
The careful analysis shows that the fusion procedure goes through
leading to {\it the same} basic functional relation (\ref{A24}).
Formula (\ref{qdet4}) for the quantum determinant also holds with
the following function $\phi (u)$:
\beq
\phi (u)=
\prod _{j=1}^{N}
\theta \left [
\begin{array}{c} \frac{1}{2} \\
\frac{1}{2}
\end{array}
\right ]
\big ( \eta (u-y_{j})\big | \tau \big ).
\label{ell4}
\eeq
We call functions of this form {\it elliptic polynomials} in $u$.
The number $N$ is {\it degree} of the elliptic polynomial.

The elliptic solution (\ref{ell1}) is no longer $GL(k)$-invariant but
is still associated with the root system $A_{k-1}$. The monodromy
matrices built out of this $R$-matrix form representations of
generalized Sklyanin algebras, 2-parameter deformations of
$U(gl_k )$ \cite{Skl}, \cite{Quano}. We refer to integrable models with
the $R$-matrix (\ref{ell1}) as {\it $A_{k-1}$-type models}.

\section{A classical view on Hirota's equation}

This section is a brief survey of HBDE in the purely classical
context. Firstly we list various types of HBDE and show their
equivalence. Then discrete versions of the zero curvature
representation and auxiliary linear problems are presented and
B\"acklund transformations are discussed.

At the first glance, the content of this section has nothing to do
with specific tools of the algebraic Bethe ansatz partially described
in the previous section. Indeed, the two sections may be read
independently of each other. However, in the next section we link
them together and show that main elements of the quantum theory
can be translated into the language of classical integrability and
embedded into the purely classical context.

\subsubsection*{Equivalent forms of the bilinear equation}

A) {\it Hirota's original form.}
In Hirota's original
notation it is
\beq
\left ( z_1 \exp (D_1 )+z_2 \exp (D_2 )+z_3 \exp (D_3 )\right )
T \cdot T =0\,,
\label{HBDE1}
\eeq
where $z_i$ are arbitrary constants, $T =T(x_1 , x_2 ,x_3 )$ is
a function of 3 variables and Hirota's $D$-operator
$D_i \equiv D_{x_i}$ is defined by
\beq
\left.\phantom{{a\over b}}
F(D_x )f(x)\cdot g(x)=F(\p _y )f(x+y)g(x-y)
\right|_{y=0 _{\phantom{ff}}}\,.
\label{D}
\eeq
In the more explicit notation eq.\,(\ref{HBDE1}) looks as follows:
\beq
z_1 T (x_1 +1)
T(x_1 -1)+
z_2 T(x_2 +1)
T(x_2 -1)+
z_3 T(x_3 +1)
T(x_3 -1)=0
\label{HBDE2a}
\eeq
(here and below we often skip variables that do not undergo shifts).

Note that the 3 variables enter in a symmetric fashion and the
equation is invariant under their permutations (and a simultaneous
permutation of $z_i$'s). The equation is also invariant under
changing the sign of any one of the variables and under the
transformation
\beq
T(x_1 , x_2 , x_3 )\rightarrow
\chi _{0}(x_1 + x_2 + x_3 )
\chi _{1}(x_2 + x_3 - x_1 )
\chi _{2}(x_1 + x_3 - x_2 )
\chi _{3}(x_1 + x_2 - x_3 )
T(x_1 , x_2 , x_3 )\,,
\label{inv1}
\eeq
where $\chi _{i}$ are arbitrary functions.

The transformation
\beq
T(x_1 , x_2 , x_3 )\rightarrow
z_{1}^{-x_{1}^{2}/2}(-z_{2})^{-x_{2}^{2}/2}z_{3}^{-x_{3}^{2}/2}
T(x_1 , x_2 , x_3 )
\label{NE1}
\eeq
converts eq.\,(\ref{HBDE2a}) into the {\it canonical form},
\beq
T(x_1 +1)T(x_1 -1)-
T(x_2 +1)T(x_2 -1)+
T(x_3 +1)T(x_3 -1)=0
\label{HBDE3}
\eeq
which does not
contain any free parameters.

After appropriate identification of the variables one recognizes the
bilinear fusion relation (\ref{A24}).
We refer to $x_1, x_2 , x_3$
as {\it direct variables} (in contrast to light cone ones, see below).

A$'$) {\it "Gauge invariant" form}:
\beq
Y(x_1 , x_2 +1, x_3 )
Y(x_1 , x_2 -1, x_3 )=\frac{
(1+Y(x_1 ,x_2 , x_3 +1))
(1+Y(x_1 ,x_2 , x_3 -1))}
{(1+Y^{-1}(x_1 +1, x_2 , x_3 ))
(1+Y^{-1}(x_1 -1 , x_2 , x_3 ))},
\label{Ysys}
\eeq
where
\beq
Y(x_1 , x_2 , x_3 )\equiv \frac{
T (x_1 , x_2 , x_3 +1)T(x_1 , x_2 , x_3 -1 )}
{T(x_1 +1, x_2 , x_3 )T(x_1 -1, x_2 , x_3 )}
\label{Y}
\eeq
is a gauge invariant quantity: the "gauge" transformation (\ref{inv1})
does not change it.

B) {\it KP-like form}. The equation for a function
$\tau (p_1 , p_2 , p_3 )$ of 3 variables reads
\beq
z_1 \tau (p_1 +1)
\tau (p_{2}+1 , p_{3} +1 )+
z_2 \tau (p_2 +1)
\tau (p_1 +1, p_{3} +1 )+
z_3 \tau (p_3 +1)
\tau (p_1 +1, p_{2}+1 )=0\,.
\label{HBDE4}
\eeq
Again, one can eliminate the arbitrary constants by the transformation
\beq
\tau (p_1 , p_2 , p_3 )\rightarrow
\left (-\frac{z_2}
{z_3}\right )^{p_1 p_2 }
\left (-\frac{z_2}
{z_1}\right )^{p_2 p_3 }
\tau (p_1 , p_2 , p_3 )\,.
\label{NE2}
\eeq

C) {\it 2DTL-like form}:
\beq
\tau _{x}(t,\bar t +1)\tau _{x}(t+1, \bar t)-
\tau _{x}(t,\bar t )\tau _{x}(t+1, \bar t +1)=
r\tau _{x+1}(t,\bar t +1)\tau _{x-1}(t+1, \bar t)\,,
\label{HBDE6}
\eeq
where $\tau _{x}(t, \bar t)$ is a function of the 3 variables and
$r$ is an arbitrary constant. The variables
$t, \bar t$ are called {\it light cone coordinates}.
Note that in this
form the permutation symmetry is lost. However, an analogue of
eq.\,(\ref{inv1}) holds true:
\beq
\tau _{x}(t,\bar t) \rightarrow
\chi _{0}(2x+2t)
\chi _{1}(2t)
\chi _{2}(2\bar t )
\chi _{3}(2x-2\bar t )
\tau _{x}(t,\bar t)
\label{inv2}
\eeq
sends solutions to solutions. The transformation
\beq
\tau _{x}(t,\bar t) \rightarrow
r^{-x^{2}/2}
\tau _{x}(t,\bar t)
\label{NE3}
\eeq
sends eq.\,(\ref{HBDE6}) to its canonical form ($r=1$).

At last we present the linear substitutions
which make canonical forms of equations a), b), c) equivalent.

A)$\leftrightarrow$ B):
$$T (x_1 , x_2 , x_3 )=
\tau (p_1 , p_2 , p_3 ),$$
\beq
x_1 =p_2 +p_3 \,,\;\;\;\;\;\;
x_2 =p_1 +p_3 \,,\;\;\;\;\;\;
x_3 =p_1 +p_2 \,;
\label{lin1}
\eeq

B)$\leftrightarrow$ C):
$$\tau (p_1 , p_2 , p_3 )=
\tau _{x}(t, \bar t),$$
\beq
p_1 =t \,,\;\;\;\;\;\;
p_2 =x-\bar t \,,\;\;\;\;\;\;
p_3 =\bar t \,;
\label{lin2}
\eeq

A)$\leftrightarrow$ C):
$$T(x_1 , x_2 , x_3 )=
\tau _{x}(t, \bar t),$$
\beq
x_1 =x \,,\;\;\;\;\;\;
x_2 =x+t-\bar t \,,\;\;\;\;\;\;
x_3 =t+\bar t \,;
\label{lin3}
\eeq

Clearly, these linear substitutions are not unique. All other
possibilities can be obtained from the given one by applying a
transformation of the form
$(x_1 , x_2 , x_3 )\rightarrow
(\pm x_{P(1)},\pm x_{P(2)},\pm x_{P(3)})$, where $P$ is a permutation.
Using formulas (\ref{lin1})-(\ref{lin3}) one can easily obtain gauge
invariant forms of equations B) and C).

\subsubsection*{Discrete Zakharov-Shabat representation}

The reformulation of classical nonlinear integrable equations as
flatness conditions for a two-dimensional connection is the basic
ingredient of the theory. The flatness means that subsequent shifts
along any pair of the time flows commute. These conditions are known as
{\it Zakharov-Shabat equations} or {\it zero curvature representation}.

Let us consider a family of difference operators acting in
the space of scalar functions of a variable $u$:
\beq
M^{(z)}=e^{\p _{u}}-z
\frac{ \tau ^{t}(u)\tau ^{t+1}(u+1) }
{ \tau ^{t+1}(u)\tau ^{t}(u+1) } ,
\label{Z1}
\eeq
\beq
\bar M^{(\bar z)}=-\bar z+
\frac{ \tau ^{\bar t}(u-1)\tau ^{\bar t+1}(u+1) }
{ \tau ^{\bar t+1}(u)\tau ^{\bar t}(u) }
e^{-\p _{u}}\,,
\label{Z2}
\eeq
where $z$, $\bar z$ are
arbitrary parameters, the shift operator
$e^{\p _{u}}$ obeys
$e^{\pm \p _{u}}f(u)=f(u\pm 1)
e^{\pm \p _{u}}$, $\tau (u)$ is a function of $u$ depending also on
a number of extra variables $t,p, \ldots ; \bar t, \bar p , \ldots $
When it is necessary to show the
dependence on these variables explicitly, we write, e.g.
$M^{(z)}(t,p, \ldots )$, etc. We call $M^{(z)}$, $\bar M^{(\bar z)}$
{\it $M$-operators};
note that their coefficients
are gauge invariant.

\begin{th}
The discrete Zakharov-Shabat equations (zero curvature conditions)
\beq
M^{(w)}(p,t+1)
M^{(z)}(p,t)=
M^{(z)}(p+1,t)
M^{(w)}(p,t)\,,
\label{Z3}
\eeq
\beq
\bar M^{(\bar w)}(\bar p,\bar t+1)
\bar M^{(\bar z)}(\bar p,\bar t)=
\bar M^{(\bar z)}(\bar p+1,\bar t)
\bar M^{(\bar w)}(\bar p,\bar t)\,,
\label{Z4}
\eeq
\beq
\bar M^{(\bar z)}(t+1,\bar t)
M^{(z)}(t,\bar t)=
M^{(z)}(t,\bar t +1)
\bar M^{(\bar z)}(t,\bar t)
\label{Z5}
\eeq
are equivalent to the following bilinear relations for
$\tau (u)=\tau ^{t,\bar t , p, \bar p , \ldots }(u)$:
\beq
z\tau ^{p+1, t}(u)
\tau ^{p, t+1}(u+1)
-w\tau ^{p, t+1}(u)
\tau ^{p+1, t}(u+1)+
H_{1}(p,t)\tau ^{p+1, t+1}(u)
\tau ^{p, t}(u+1)=0\,,
\label{Z8}
\eeq
\beq
\bar z \tau ^{\bar p+1, \bar t}(u+1)
\tau ^{\bar p, \bar t+1}(u)
-\bar w \tau ^{\bar p, \bar t+1}(u+1)
\tau ^{\bar p+1, \bar t}(u)+
H_{2}(\bar p,\bar t)\tau ^{\bar p+1, \bar t+1}(u+1)
\tau ^{\bar p, \bar t}(u)=0\,,
\label{Z8a}
\eeq
\beq
z\tau ^{t, \bar t+1}(u)
\tau ^{t+1, \bar t}(u)+
H_{3}(t, \bar t )\tau ^{t, \bar t}(u)
\tau ^{t+1, \bar t +1}(u)=
(\bar z )^{-1}\tau ^{t, \bar t+1}(u+1)
\tau ^{t+1, \bar t}(u-1)\,,
\label{Z8b}
\eeq
respectively, with arbitrary $u$-independent functions $H_i$.
\end{th}
The proof consists in straightforward commutation of the $M$-operators.

We see that

eq.\,(\ref{Z8}) coincides with HBDE in the
KP-like form (\ref{HBDE4}) under the identification
$\tau ^{p,t}(u)=\tau (p,t,u )$ and
$$
z=z_1 , \;\;\;\;\;\; w=-z_2 , \;\;\;\;\;\; H_1 =z_3\,;
$$

eq.\,(\ref{Z8a}) coincides with HBDE in the
KP-like form (\ref{HBDE4}) under the identification
$\tau ^{\bar p,\bar t}(u)=\tau (\bar p,\bar t,-u)$ and
$$
\bar z=z_1 , \;\;\;\;\;\; \bar w=-z_2 , \;\;\;\;\;\; H_2 =z_3\,;
$$

eq.\,(\ref{Z8b}) coincides with HBDE in the
2DTL-like form (\ref{HBDE6}) under the identification
$\tau ^{t,\bar t}(u)=\tau _{u}(t,\bar t)$ and
$$
z\bar z =r^{-1}\,, \;\;\;\;\;\;  H_3 =-z\,.
$$

The variables $t,\bar t$ provide a family of commuting flows
parametrized by $z, \bar z$. In fact one can realize $M$-operators
as difference operators in any one of them. The variable in which
$M$-operators act will be called {\it the reference variable}.
The functions $H_i$
are automatically fixed to
be the constants if one requires the simultaneous compatibility
of all the corresponding zero curvature conditions.

\subsubsection*{Linearization of HBDE}

The zero curvature conditions are equivalent to compatibility
of an overdetermined system of linear difference equations.
These linear equations are called {\it auxiliary
linear problems} (ALP). They play a very important role in the theory.
Common solutions to ALP (called wave functions) carry the complete
information about solutions to the nonlinear equations. All the
properties of the latter can be translated into the language of the
ALP. This is what we mean by the linearization of HBDE.

The discrete Zakharov-Shabat equations (\ref{Z3})-(\ref{Z5}) imply
the compatibility of the linear problems
\beq
M^{(z)}\psi ^{t, \bar t}(u)=
-z\psi ^{t+1, \bar t}(u)\,,
\label{S1}
\eeq
\beq
\bar M^{(\bar z)}\psi ^{t, \bar t}(u)=
-\bar z\psi ^{t, \bar t +1}(u)
\label{S2}
\eeq
for any discrete flows labeled by $z, \bar z$.
Note that the "eigenvalues" in the r.h.s. can be arbitrary: this is
the matter of redefinition of $\psi ^{t,\bar t}(u)$. Our choice
corresponds to the smooth continuum limit.

More explicitly, eqs.\,(\ref{S1}), (\ref{S2}) read
(see (\ref{Z1}), (\ref{Z2})):
\beq
zV^{t,\bar t}(u)\psi ^{t,\bar t}(u)-
\psi ^{t,\bar t}(u+1)=z\psi ^{t+1, \bar t}(u)\,,
\label{S3}
\eeq
\beq
\bar z
\psi ^{t,\bar t}(u)-C^{t,\bar t}(u)
\psi ^{t,\bar t}(u-1)=\bar z \psi ^{t, \bar t +1}(u)\,,
\label{S4}
\eeq
where
\beq
V^{t,\bar t}(u)=
\frac{ \tau ^{t,\bar t}(u)\tau ^{t+1, \bar t}(u+1)}
{\tau ^{t+1,\bar t}(u)\tau ^{t, \bar t}(u+1)},
\label{S5}
\eeq
\beq
C^{t,\bar t}(u)=
\frac{ \tau ^{t,\bar t}(u-1)\tau ^{t, \bar t +1}(u+1)}
{\tau ^{t,\bar t +1}(u)\tau ^{t, \bar t}(u)}.
\label{S6}
\eeq

These formulas become more symmetric in terms of the "unnormalized"
wave function
\beq
\rho (u)=\psi (u)\tau (u)\,.
\label{S7}
\eeq
Making this substitution in (\ref{S1}), (\ref{S4}), we get:
\beq
z\tau ^{t+1, \bar t}(u+1)
\rho ^{t,\bar t}(u)-
\tau ^{t+1, \bar t}(u)
\rho ^{t,\bar t}(u+1)=
z\tau ^{t, \bar t}(u+1)
\rho ^{t+1,\bar t}(u)\,,
\label{S8}
\eeq
\beq
\bar z \tau ^{t, \bar t +1}(u)
\rho ^{t,\bar t}(u)-
\tau ^{t, \bar t +1}(u+1)
\rho ^{t,\bar t}(u-1)=
\bar z\tau ^{t, \bar t}(u)
\rho ^{t,\bar t +1}(u)\,.
\label{S9}
\eeq

\subsubsection*{Duality}

The ALP (\ref{S8}), (\ref{S9}) have a remarkable property:
they are symmetric under interchanging
$\tau$ and $\rho$. Furthermore, one may treat them as linear problems
for the function $\tau$, the compatibility condition being a
nonlinear equation for $\rho$. This equation is again HBDE. We refer
to this fact as the {\it duality} between "potentials" $\tau$
and "wave functions" $\rho$.

More precisely, rewriting eqs.\,(\ref{S3}), (\ref{S4}) as linear
equations for
$$
\tilde \psi ^{t,\bar t}(u)=
\frac{ \tau ^{t+1, \bar t+1}(u+1)}
{ \rho ^{t+1, \bar t+1}(u+1)} =
\left ( \psi ^{t+1, \bar t+1}(u+1)\right ) ^{-1},
$$
we have
\beq
\left ( e^{-\p _{u}}-z)
\tilde V^{t,\bar t}(u)\right )\tilde \psi ^{t,\bar t}(u)=
-z\tilde \psi ^{t-1, \bar t}(u)\,,
\label{D1}
\eeq
\beq
\left (\bar z-
\tilde C^{t,\bar t}(u+1)e^{\p _{u}} \right )\tilde \psi ^{t,\bar t}(u)=
\bar z\tilde \psi ^{t, \bar t -1}(u)\,,
\label{D2}
\eeq
where $\tilde V$ and $\tilde C$ are given by the same formulas
(\ref{S5}), (\ref{S6}) with $\rho$ in place of $\tau$. After the
replacing $\rho \rightarrow \tau$ the difference operators in the
l.h.s. become formally adjoint to the operators (\ref{Z1}), (\ref{Z2})
(defined by the rule $(f(u)e^{k\p _{u}})^{\dag}=e^{-k\p _{u}}f(u)$).
It then follows that the compatibility conditions are described by
Theorem 3.1 with $\tau$ replaced by $\rho$.

Therefore, passing from a given solution $\tau$ to $\rho$ we have got
a new solution to HBDE. This is a B\"acklund-type transformation.

\subsubsection*{B\"acklund flows}

One may repeat the procedure
described above once again starting from $\rho$ and,
moreover, consider a chain of successive transformations of this
kind. Let us introduce an additional variable $m$ to mark
steps of the "flow" along this chain and let $\tau ^{t,\bar t}_{m}(u)$ ,
$\rho ^{t,\bar t}_{m}(u)$ be $\tau$ and $\rho$ at $m$-th step.

The B\"acklund flow is
defined by
\beq
\rho ^{t,\bar t}_{m}(u)= z^{u}
\tau ^{t,\bar t}_{m+1}(u)\,.
\label{D3a}
\eeq
This means that $\tau$ at the next step of the "B\"acklund time" $m$
is put equal to a solution $\rho$ of the linear equations (\ref{S8}),
(\ref{S9}) (up to the factor $z^{u}$). Then these linear
problems become {\it bilinear equations} for $\tau _{m}$:
\beq
\tau _{m} ^{t+1}(u)\tau _{m+1}^{t}(u+1)-
\tau _{m} ^{t+1}(u+1)\tau _{m+1}^{t}(u)+
\tau _{m} ^{t}(u+1)\tau _{m+1}^{t+1}(u)=0\,,
\label{D4}
\eeq
\beq
\tau _{m} ^{\bar t+1}(u)\tau _{m+1}^{\bar t}(u)-
\tau _{m} ^{\bar t}(u)\tau _{m+1}^{\bar t+1}(u)=
(z\bar z)^{-1}
\tau _{m} ^{\bar t+1}(u+1)\tau _{m+1}^{\bar t}(u-1)\,,
\label{D5}
\eeq
where $\bar t$ (resp., $t$) in eq.\,(\ref{D4}) (resp., (\ref{D5}))
is skipped.

Similarly, defining the second B\"acklund flow (the B\"acklund time is
now denoted by $\bar m$),
\beq
\rho ^{t,\bar t}_{\bar m}(u)= (\bar z)^{-u}
\tau ^{t,\bar t}_{\bar m+1}(u+1)\,,
\label{D3b}
\eeq
we get from (\ref{S8}), (\ref{S9}):
\beq
\tau _{\bar m} ^{t+1}(u)\tau _{\bar m+1}^{t}(u)-
\tau _{\bar m} ^{t}(u)\tau _{\bar m+1}^{t+1}(u)=
(z\bar z)^{-1}
\tau _{\bar m+1} ^{t}(u+1)\tau _{\bar m}^{t+1}(u-1)\,,
\label{D6}
\eeq
\beq
\tau _{\bar m} ^{\bar t}(u)\tau _{\bar m+1}^{\bar t+1}(u+1)-
\tau _{\bar m} ^{\bar t+1}(u)\tau _{\bar m+1}^{\bar t}(u+1)+
\tau _{\bar m} ^{\bar t+1}(u+1)\tau _{\bar m+1}^{\bar t}(u)=0\,.
\label{D7}
\eeq
In these equations one immediately recognizes different forms
of HBDE described by Theorem 3.1. So, we conclude that the
B\"acklund flows can be identified with the
commuting discrete flows.

\subsection*{Comments and references}

1. The bilinear difference equation was suggested by R.Hirota
in the paper \cite{Hirota1} which summarized his earlier
studies on discretization of nonlinear integrable equations
\cite{HirotaKdV}-\cite{Hirota5}. HBDE (\ref{HBDE3}) can be
viewed as an integrable discrete analogue of the 2-dimensional
Toda lattice.

2. As it was first noticed by T.Miwa \cite{Miwa1}, discrete Hirota's
equations can be obtained from the continuous KP hierarchy by choosing
the time flows to be certain infinite
combinations of the standard continuous flows
of the hierarchy (Miwa's transformation).
This approach was further developed in the papers
\cite{Miwa2},\,\cite{Miwa3}.
The different forms of HBDE arise when one applies Miwa's
transformation to different continuous hierarchies.
The function $\tau _{u} (t, \bar t)$ can be identified with
the $\tau$-function of the hierarchies (restricted to a finite
number of the discrete flows).

3. The "gauge invariant" form
(\ref{Ysys}) of HBDE is a discrete counterpart of nonlinear
integrable equations written in terms of potentials and fields
rather than $\tau$-functions. Eq.\,(\ref{Y}) is a discrete version of the
famous formula $U(x)=2\p _{x}^{2}\log \tau (x)$.
Some particular cases of equation (\ref{Ysys}) emerge naturally in
thermodynamic Bethe ansatz \cite{Zam}-\cite{BLZ}.

4. An example of the
discretized zero curvature representation for
eq.\,(\ref{HBDE3}) was given by R.Hirota \cite{Hirota1}.
In the physical language, the discrete connection is a
lattice gauge field. The approach emphasizing the relation to gauge field
theories on the lattice was developed by S.Saito and N.Saitoh \cite{SS}.
We have presented
these results in a modified form which makes the theory
completely parallel to the 2DTL theory.
The bilinear form of B\"acklund transformations was discussed by
R.Hirota \cite{Hirota4}. The $\tau \leftrightarrow \rho$ duality was
pointed out in ref.\,\cite{SS}.

5. Similarly to the 2DTL hierarchy, there exists an infinite
hierarchy of bilinear difference equations. For the explicit
form of higher equations of the hierarchy see ref.\,\cite{OHTI}.
We remark
that the general quantum fusion rules in the bilinear form
(see e.g. eq.\,(\ref{A25})) coincide with the higher HBDE-like
equations (after a linear change of variables).

6. In the continuum limit ($z \rightarrow \infty$,
$\bar z \rightarrow \infty$) eqs.\,(\ref{S3}), (\ref{S4}) turn
into the familiar ALP for the 2DTL \cite{UT}:
\begin{eqnarray}
&&\p _{t}\psi _{n}=\psi _{n+1}+\p _{t}(\log \frac{\tau _{n+1}}
{\tau _{n}})\psi _{n}\,,
\nonumber \\
&&\p _{\bar t}\psi _{n}=\frac{\tau _{n+1}\tau _{n-1}}
{\tau _{n}^{2}}\psi _{n-1}\,,
\label{S10}
\end{eqnarray}
where $n$ is identified with $u$. The compatibility condition is
\beq
\p_{t}\tau _{n}\p_{\bar t}\tau _{n}-\tau _{n}\p_{t}\p_{\bar t}\tau _{n}=
\tau _{n+1}\tau _{n-1}\,,
\label{toda1}
\eeq
which in terms of
$$\varphi _{n}(t, \bar t)=\log \frac{\tau _{n+1}(t, \bar t)}
{\tau _{n}(t, \bar t)}$$
acquires the form
\beq
\p_{t}\p_{\bar t}\varphi _{n}=e^{\varphi _{n}-\varphi _{n-1}}-
e^{\varphi _{n+1}-\varphi _{n}}\,,
\label{toda2}
\eeq
the first non-trivial equation of the 2DTL.

\section{Bethe ansatz results from the Hirota equation}

We are going to show how to reproduce the quantum Bethe ansatz
results by solving the classical Hirota difference equation
(\ref{A24}):
\beq
T^{a}_{s}(u+1)T^{a}_{s}(u-1)-
T^{a}_{s+1}(u)T^{a}_{s-1}(u)=
T^{a+1}_{s}(u)T^{a-1}_{s}(u)\,.
\label{A24a}
\eeq
In general, HBDE has many solutions of very different
nature. To extract the solutions of interest, we should first of all
specify boundary and analytic conditions for $T_{s}^{a}(u)$.

\subsubsection*{Boundary conditions in $a$ and $s$}

The values of
$T_{s}^{0}(u)$ and $T_{s}^{k}(u)$ for $A_{k-1}$-type models
should be considered as boundary conditions (b.c.). Recall that
$T_{s}^{0}(u)=1$ and
$T_{s}^{k}(u)$ is the quantum determinant given
explicitly by eq.\,(\ref{qdet4}):
\beq
T_{s}^{k}(u)=\prod _{p=0}^{s-1}\left (
\phi (u-s+k+2+2p)
\prod _{l=1}^{k-1}
\phi (u-s-k+2p+2l)\right ).
\label{qdet4a}
\eeq
Let us note that
$T_{s}^{k}(u)$
obeys the discrete d'Alembert equation:
\beq
T_{s}^{k}(u+1)T_{s}^{k}(u-1)=T_{s+1}^{k}(u)T_{s-1}^{k}(u).
\label{B1}
\eeq
This is easily seen from eq.\,(\ref{qdet4a}). To put it differently,
recall that the general solution to the discrete d'Alembert equation
is $\chi _{+}(u+s)\chi _{-}(u-s)$, i.e. it is factorized into a
product of a "holomorphic" function $\chi _{+}$ and an
"antiholomorphic" function $\chi _{-}$. It is possible to represent
$T_{s}^{k}(u)$ in this form indeed. For instance, in the rational
case we have:
\begin{eqnarray}
\label{qdet4b}
&&T_{s}^{k}(u)=\frac{\chi _{k}(u+s)}
{\chi _{k}(u-s)}\,,\nonumber \\
&&\chi _{k}(u)=2^{Nku/2}\prod _{i=1}^{N}
\left ( \Gamma \big ( \frac{u+k-y_i }{2}+1\big )
\prod _{l=1}^{k-1}
\Gamma \big ( \frac{u-k-y_i }{2}+l\big )\right ),
\end{eqnarray}
where $\Gamma (u)$ is the gamma-function.
Combining this property with the Hirota equation, we arrive at
the following b.c.:
\beq
T^{a}_{s}(u)=0 \;\;\;\;\;\mbox{as}\;\;a<0 \;\;\;\mbox{or}\;\;\; a>k.
\label{B2}
\eeq

The b.c. in $s$ reads
\beq
T_{s}^{a}(u)=0
\;\;\;\;\;\mbox{as}\;\;\; -k<s<0,\;\;\;\mbox{and}
\;\;\;0<a<k\,. \label{B3} \eeq
The meaning of this condition will be discussed later.

\subsubsection*{Analytic conditions in $u$}

A very important condition in models on finite lattices
(which follows, eventually, from the Yang-Baxter
equation) is that $T^{a}_{s}(u)$ for any fixed
$a,s,u$ has to be an
elliptic polynomial in the spectral parameter $u$ (see (\ref{ell4})
multiplied possibly by an exponential function.

More precisely, the gauge invariant quantity
\beq
\label{B4}
Y^a_s(u)=\frac{T_{s+1}^{a}(u)T_{s-1}^{a}(u)}
{T_{s}^{a+1}(u)T_{s}^{a-1}(u)}
\eeq
has to be an
elliptic
function of $u$ having $2N$ zeros and
$2N$ poles in the fundamental domain.
This implies that $T_{s}^{a}(u)$ has the general form
\beq
T_{s}^{a}(u)=A_{s}^{a}e^{\mu (a,s)u}
\prod _{j=1}^{N(a,s)} \theta
\left [ \begin{array}{c}1/2 \\ 1/2 \end{array} \right ]
\big (\eta (u-w_{j}^{(a,s)})\big )\,,
\label{B5}
\eeq
where $w_{j}^{(a,s)}$, $A_{s}^{a}$, $\mu (a,s)$ do not depend on $u$
(and meet some additional requirements). For models with rational
$R$-matrices $T_{s}^{a}(u)$ degenerates to a usual polynomial
(multiplied by the exponential function).
We note that the b.c. (\ref{B2}), (\ref{B3}) are consistent with the
required analytic conditions.

\subsubsection*{Normalizations}

By normalization we mean multiplying $T_{s}^{a}(u)$ by the "gauge"
factors
$\chi _{0} (u+s+a)
\chi _{1} (u+s-a)
\chi _{2} (u-s+a)
\chi _{3} (u-s-a)$. Boundary and analytic conditions may depend on the
gauge. We point out the following three distinguished normalizations.

i) {\it Det-normalization} was already discussed in detail:
$T_{s}^{0}(u)=1$, $T_{s}^{k}(u)$ is the quantum determinant
(\ref{qdet4a}).

ii) {\it Minimal polynomial (MP) normalization}.
The"gauge" invariance (\ref{inv1}) allows one to remove all zeros
from the characteristics $a\pm s\pm u =\mbox{const}$. These are just
the "trivial" zeros brought by the fusion procedure. The minimal
polynomial appears in the gauge (\ref{A20}).
The boundary values at $a=0,k$ then become:
\begin{eqnarray}
T^{0}_{s}(u)&=&\phi (u-s),\nonumber\\
T^{k}_{s}(u)&=&\phi (u+s+k)\,.
\label{B6}
\end{eqnarray}
Note that in this normalization $T^0$ is an "antiholomorphic" function
(i.e. depends on $u+s$ only) while
$T^k$ is a "holomorphic" function
(i.e. depends on $u-s$ only).
The most important feature of the MP normalization is that the
elliptic polynomials $T_{s}^{a}(u)$ have one and the same degree $N$
for all values $a,s$ (when
$T_{s}^{a}(u)$ is not identically zero).
In the sequel, we use the MP normalization.

iii) {\it Canonical normalization}. Since the functions $T_{s}^{0}(u)$
and $T_{s}^{k}(u)$
at the "boundaries" obey the discrete d'Alembert equation,
they can be gauged away, that is can be made equal to unity:
$T_{s}^{0}(u)=
T_{s}^{k}(u)=1$. This allows one to simplify the equations for the
price of imposing much more complicated analytic properties of solutions.
Here we will not discuss this normalization.

\subsubsection*{Zero curvature representation}

Let us represent the bilinear fusion relation (\ref{A24a}) as a
discrete zero curvature condition. To do that, we can make use
of Theorem 2.1. Consider eq.\,(\ref{Z8b}) and identify
\beq
T_{s}^{a}(u)=
\tau ^{t, \bar t}(u),
\label{B7}
\eeq
where
\beq
s=-t-\bar t,\;\;\;\;\;\;\; a=u+t-\bar t
\label{B8}
\eeq
(cf.\,(\ref{lin3})). We have to substitute
$\p _{t}
\rightarrow -\p _{s}+\p _{a}$,
$\p _{\bar t}
\rightarrow -\p _{s}-\p _{a}$ and
$$
\left.\phantom{{a\over b}}
\p _{u}
\right|_{t,\bar t =\mbox{const}_{\phantom{ff}}}
\rightarrow
\left.\phantom{{a\over b}}
\p _{u}
\right|_{a,s=\mbox{const}_{\phantom{ff}}}
+\p _{a},
$$
so the $M$-operators read
\beq
M(a,s)=e^{\p _{u}+\p _{a}}
-z
\frac{T_{s}^{a}(u)T_{s-1}^{a+2}(u+1)}
{T_{s-1}^{a+1}(u)T_{s}^{a+1}(u+1)},
\label{B9}
\eeq
\beq
\bar M(a,s)=
-\bar z+
\frac{T_{s}^{a-1}(u-1)T_{s-1}^{a}(u+1)}
{T_{s-1}^{a-1}(u)T_{s}^{a}(u)}
e^{-\p _{u}-\p _{a}}.
\label{B10}
\eeq
In terms of the operators
\beq
{\cal M}=
e^{\p _{s}-\p _{a}}M(a,s), \;\;\;\;\;\;\;
\bar {\cal M}=
e^{\p _{s}+\p _{a}}\bar M(a,s)
\label{B11}
\eeq
the Zakharov-Shabat equation (\ref{Z5}) acquires the form of a
commutativity condition:
\beq
\phantom{a} [{\cal M},\, \bar {\cal M}]=0\,.
\label{B12}
\eeq

\subsubsection*{Linear problems}

The commutativity of two operators implies the existence of a
common eigenfunction:
\beq
{\cal M}\Psi ^{a,s}(u)=E
\Psi ^{a,s}(u),
\;\;\;\;\;\;\;
\bar {\cal M}\Psi ^{a,s}(u)=\bar E
\Psi ^{a,s}(u).
\label{B13}
\eeq
Explicitly, these equations read
\beq
\Psi ^{a,s}(u+1)
-z
\frac{T_{s}^{a-1}(u)T_{s-1}^{a+1}(u+1)}
{T_{s-1}^{a}(u)T_{s}^{a}(u+1)}
\Psi ^{a-1, s}(u)=E
\Psi ^{a, s-1}(u),
\label{B14}
\eeq
\beq
-\bar z \Psi ^{a,s}(u)+
\frac{T_{s}^{a-1}(u-1)T_{s-1}^{a}(u+1)}
{T_{s-1}^{a-1}(u)T_{s}^{a}(u)}
\Psi ^{a-1, s}(u-1)=\bar E
\Psi ^{a-1, s-1}(u).
\label{B15}
\eeq
Passing to the "unnormalized" wave function,
\beq
F_{s}^{a}(u)=
T_{s}^{a}(u)\Psi ^{a,s}(u),
\label{B16}
\eeq
we get
\begin{eqnarray}
&&T_{s-1}^{a}(u)F_{s}^{a}(u+1)-
zT_{s-1}^{a+1}(u+1)F_{s}^{a-1}(u)= E
T_{s}^{a}(u+1)F_{s-1}^{a}(u)\,, \nonumber\\
&&T_{s-1}^{a}(u+1)F_{s}^{a-1}(u-1)-\bar z
T_{s-1}^{a-1}(u)F_{s}^{a}(u)=\bar E
T_{s}^{a}(u)F_{s-1}^{a-1}(u)\,.
\label{B17}
\end{eqnarray}

Now, to identify eq.\,(\ref{Z8b}) with
eq.\,(\ref{A24a}) literally, we set
$z=\bar z =H_3 =-1$. Besides, redefining
$\Psi ^{a,s}(u)
\rightarrow E^{(s-a)/2}(\bar E)^{(s+a)/2}
\Psi ^{a,s}(u)$, we can always choose $E=\bar E =1$ without
loss of generality. In this way we get the following ALP:
\begin{eqnarray}
&&T_{s+1}^{a+1}(u)F_{s}^{a}(u)-
T_{s}^{a+1}(u+1)F_{s+1}^{a}(u-1)=
T_{s}^{a}(u)F_{s+1}^{a+1}(u)\,, \nonumber\\
&&T_{s+1}^{a}(u+1)F_{s}^{a}(u)-
T_{s}^{a}(u)F_{s+1}^{a}(u+1)=
T_{s}^{a+1}(u+1)F_{s+1}^{a-1}(u)\,.
\label{B18}
\end{eqnarray}

An advantage of the light cone coordinates
$t, \bar t$ is that they are
separated in the linear problems (compare (\ref{S8}), (\ref{S9}) with
(\ref{B17})). However,
in contrast to $a,s,u$ they do not have an immediate physical meaning.

Due to the duality property (Sect.\,3) $F_{s}^{a}(u)$ obeys the same
HBDE:
\beq
F^{a}_{s}(u+1)F^{a}_{s}(u-1)-
F^{a}_{s+1}(u)F^{a}_{s-1}(u)=
F^{a+1}_{s}(u)F^{a-1}_{s}(u)\,.
\label{B19}
\eeq
We require $F_{s}^{a}(u)$ to have the same analytic properties as
$T_{s}^{a}(u)$ (though, degree of the elliptic polynomial may be
different).

The b.c.
(\ref{B2}) allows one to impose a similar condition for $F$:
\beq
F^{a}_{s}(u)=0 \;\;\;\;\; \mbox{as}\;\; a<0 \;\;\; \mbox{or}\;\;\; a>k-1
\label{B20}
\eeq
so that the number of non-zero functions $F$ is one less than the number of
$T$'s. The functions $F^a$ at the ends of the Dynkin graph ($a=0, k-1$)
have a very special form.
From the second equation of the pair
(\ref{B18}) at $a=0$ and from
the first one at $a=k-1$ it follows that $F^{0}_{s}(u)$
(respectively, $F^{k-1}_{s}(u)$)
depends on one light cone variable $u-s$ (resp., $u+s$). We introduce
a special notation for them:
\beq
F^{0}_{s}(u)=Q_{k-1}(u-s),\;\;\;\;\;F^{k-1}_{s}(u)
={\bar Q}_{k-1}(u+s).
\label{B21}
\eeq
Furthermore, it can be shown that
the important condition (\ref{B6}) relates the
functions $Q$ and $\bar Q$ as follows:
\beq
{\bar Q}_{k-1}(u)=Q_{k-1}(u+k-1).
\label{B22}
\eeq

Therefore, the analytic properties and b.c. for
$F^{a}_{s}(u)$ are the same as for $T_{s}^{a}(u)$ under a substitution
$\phi (u)$ by
$Q_{k-1}(u)$. The only change is a reduction of the Dynkin graph:
$k\rightarrow k-1$. Using this property, one can successively reduce the
$A_{k-1}$-problem up to $A_1$. Below we use this trick  to derive
$A_{k-1}$ ("nested") Bethe ansatz equations.

\subsubsection*{Nested Bethe ansatz and B\"acklund flows}

Quantum integrable models with internal degrees of freedom can be solved
by the nested (hierarchical) Bethe ansatz method. The method consists
essentially in integration over a part of degrees of freedom by an ansatz
of Bethe type, the effective hamiltonian being again integrable.
Repeating this step several times, one reduces the model to an integrable
model without internal degrees of freedom which is solved by the usual
Bethe ansatz.

The classical face of this scheme is a chain of B\"acklund transformations,
i.e. passing from solutions of the non-linear equation to (properly
normalized) solutions of the auxiliary linear problems discussed
in Sect.\,3.

To elaborate the chain of these transformations, let
$m=0,\, 1,\ldots ,k$ mark steps of the flow
$A_{k-1}\rightarrow A_1$
and let
$F^{a,m}_{s}(u)$ be a solution to the linear problem at $(k-m)$-th
level. In this notation, $F^{a,k}_{s}(u)=T_{s}^{a}(u)$ and
$F^{a,k-1}_{s}(u)=F^{a}_{s}(u)$
is the
corresponding wave function.
For each level $m$ the function
$F^{a,m}_{s}(u)$ obeys HBDE of the form (\ref{B19}) with the b.c.
\beq \label{B23}
F^{a,m}_{s}(u)=0 \;\;\;\;\; \mbox{as}\;\; a<0 \;\;\;
 \mbox{or}\;\;\; a>m\,.
\eeq
The first ($a=0$) and the last ($a=k-1$) components of the vector
$F^{a,m}_{s}(u)$ obey the discrete d'Alembert
equation and under the
condition
(\ref{B21}) are "antiholomorphic" and "holomorphic" functions
respectively. We
denote them as follows:
\beq
F^{0,m}_{s}(u)\equiv Q_m (u-s)\,, \;\;\;\;\;\;
F^{m,m}_{s}(u)\equiv  {\bar Q}_m (u+s)\,,
\label{B24}
\eeq
where it is implied that $Q_k (u)=\phi (u)$.
Furthermore, it can be shown that
the relation
\beq
\bar Q_m(u)=Q_{m}(u+m)
\label{B25}
\eeq
can be imposed simultaneously for all $1\leq m \leq k$.

The linear problems (\ref{B18}) at level
$m$,
\beq F^{a+1,m+1}_{s+1}(u)F^{a,m}_{s}(u)-
F^{a+1,m+1}_{s}(u+1)F^{a,m}_{s+1}(u-1)=
F^{a,m+1}_{s}(u)F^{a+1,m}_{s+1}(u)\,,
\label{B26}
\eeq
\beq
F^{a,m+1}_{s+1}(u+1)F^{a,m}_{s}(u)-
F^{a,m+1}_{s}(u)F^{a,m}_{s+1}(u+1)=
F^{a+1,m+1}_{s}(u+1)F^{a-1,m}_{s+1}(u)
\label{B27}
\eeq
look as bilinear equations  for a functions of 4 variables.
However, eq. (\ref{B26}) (resp., eq. (\ref{B27}))  leaves
the hyperplane $u+s-a=\mbox{const}$ (resp.,
$u-s-a=\mbox{const}$) invariant, and actually depends on three variables.

Restricting the variables in eq. (\ref{B26}) to the hyperplane
$u+s-a=0$, i.e. setting
\beq
\tau _{u}(t,a)\equiv F^{a,k-m}_{a-u}(u)
\label{B28}
\eeq
we reduce it to the 2DTL-like form of HBDE (\ref{HBDE6})
for $\tau _{u}(m,a)$ with
$m$ and $a$ being the light cone coordinates.
The b.c. is
\beq
\tau _{u}(m,0)=Q_{k-m}(2u),
\;\;\;\; \tau _{u}(m,k-m)=\bar Q_{k-m}(m-k)=\mbox{const}.
\label{B29}
\eeq
A similar equation can be obtained from the second
linear problem (\ref{B27}). They are nothing else than eqs.\,(\ref{D5}),
(\ref{D7}) for the B\"acklund flows.

It is convenient to
visualize this array of
$\tau$-functions on a diagram; here is an example for the $A_3$-case ($k=4$):
\beq
\begin{array}{ccccccccccccc}
0&&1&&0&&&&&&&&\\
&&&&&&&&&&&&\\
0&&Q_1 (u-s)&& \bar Q_1 (u+s)&&0&&&&&&\\
&&&&&&&&&&&&\\
0&&Q_2 (u-s)&&F_{s}^{1,2}(u)&& \bar Q_2 (u+s)&&0&&&&\\
&&&&&&&&&&&&\\
0&&Q_3 (u-s)&&F_{s}^{1,3}(u)&&F_{s}^{2,3}(u)&& \bar Q_3 (u+s)&&0&&\\
&&&&&&&&&&&&\\
0&& Q_4 (u-s)&&T_s^{1}(u)&&T_s^{2}(u)&&T_s^{3}(u)&&
 \bar Q_4 (u+s)&&0
\end{array}
\label{diag}
\eeq
Functions in each horizontal (constant $m$) slice satisfy HBDE,
whereas functions
on the $u-s-a=\mbox{const}$ slice satisfy HBDE with $m$, $a$
being the light cone variables.

In the nested Bethe ansatz approach,
the functions $F_{s}^{a,m}(u)$ are auxiliary
objects -- eigenvalues of transfer matrices for "intermediate" models
arising at
$(k-m)$-th step (level) of the hierarchical
Bethe ansatz\footnote{Unlike eq.\,(\ref{A24a}), eqs.\,(\ref{B26}),
(\ref{B27}) can not be understood in the operator sense since $F^m$
and $F^{m+1}$ are eigenvalues of operators acting in different
quantum spaces.}.
The functions
$Q_m (u)$ play a distinguished role. They can be identified with
generalized Baxter's $Q$-operators in the diagonal representation
(see below). In the general solution to HBDE these functions are
arbitrary functional parameters. The additional requirement of
ellipticity determines them through the Bethe equations for their
zeros.

\subsubsection*{Bethe equations as a discrete dynamical system}

Recall that
the function $\tau _{u}(m,a)=F^{a,k-m}_{a-u}(u)$ (\ref{B28})
obeys HBDE in light cone
variables:
\beq
\tau _{u}(m+1,a) \tau _{u}(m, a+1)-
\tau _{u}(m,a) \tau _{u}(m+1, a+1)=
\tau _{u-1}(m+1,a) \tau _{u+1}(m, a+1)\,.
\label{B30}
\eeq
Since $\tau _{u}(m,0)=Q_{k-m}(2u)$, nested Bethe ansatz
equations can be understood as "equations of motions" for zeros
of $Q_{m}(u)$ in discrete time $m$ (level of the Bethe ansatz).
The simplest way to derive them is to consider the
auxiliary linear problems
for eq. (\ref{B30}). Here we present an example of this derivation
in the simplest possible form.

Let us assume that $Q_{m}(u)$ has the form
\beq
Q_{m}(u)=e^{\nu _{m}\eta u}\prod _{j=1}^{M_{m}}
\theta \left [ \begin{array}{c}\small{1/2}\\
\small{1/2}\end{array} \right ]
(\eta (u-u_{j}^{m}))
\label{B31}
\eeq
(note that we allow the number of roots $M_m$ to depend on $m$).
Since we are interested in dynamics in $m$ at a fixed $a$, it is
sufficient to consider only the first linear equation
(\ref{S8}) substituting $\tau ^{t,\bar t}(u)\rightarrow \tau _{u}(m,a)$.
An elementary way to derive equations of motion for roots of
$\tau _{u}(m,0)$ is to put $u$ equal to the roots of
$\rho ^{m, 0}(u)$,
$\rho ^{m,0}(u+1)$ and $\rho ^{m+1, 0}(u)$, so that
only two terms in (\ref{S8}) would survive. Combining relations
obtained in this way, one can eliminate $\rho$'s and obtain
the system of equations
\beq
\frac
{Q_{m-1}(u_{j}^{m}+2)Q_m (u_{j}^{m}-2)Q_{m+1}(u_{j}^{m})}
{Q_{m-1}(u_{j}^{m})Q_m (u_{j}^{m}+2)Q_{m+1}(u_{j}^{m}-2)}=-1\,.
\label{NBE}
\eeq
as the
necessary conditions for solutions of the form (\ref{B31})
to exist.
With the "boundary conditions"
\beq
Q_{0}(u)=1,\;\;\;\;\;\; Q_k (u)=\phi (u),
\label{B32}
\eeq
this system of $M_1 +M_2 +\ldots +M_{k-1}$ equations is equivalent to
the nested Bethe ansatz equations for $A_{k-1}$-type
quantum integrable models with Belavin's elliptic $R$-matrix.
The same
equations can be obtained for the right edge of the diagram (\ref{diag})
from the second linear equation.
In what follows we explicitly identify
our $Q$'s with similar objects known from the Bethe ansatz solution.

Let us remark that the origin of equations (\ref{NBE}) suggests to call
them as
equations of motion for the elliptic Ruijsenaars-Schneider (RS)
model in discrete time. Taking the continuum limit in $m$ (provided
$M_m=M$ does not depend on $m$), one can check that eqs. (\ref{NBE})
do yield the equations of motion for the elliptic RS model
with $M$
particles. The additional limiting procedure $\eta \rightarrow 0$ with
finite $\eta u_j =x_j$ yields the well known equations of motion for
the elliptic Calogero-Moser system of particles.

However, integrable systems of particles in discrete time have a
richer structure than their continuous counterparts. In particular,
the total number of particles may depend on the discrete time. Such a
phenomenon is possible in continuous time models only for singular
solutions, when particles can move to infinity or merge to another
within a finite period of time. Remarkably, this appears to be the
case for the solutions to eq.\,(\ref{NBE}) corresponding to
eigenstates of quantum models. It is known that the number of
excitations at $m$-th level of the nested Bethe ansatz solution
does depend on $m$. In other words, the number of "particles" in
the associated discrete time RS system is not conserved. At the
same time the numbers $M_m$ may not be arbitrary. It can be shown
that for models with elliptic $R$-matrices in case of general
position $M_m = (N/k)m$, where $N$ is the number of sites of the
lattice (degree of the elliptic polynomial $\phi (u)$).
In trigonometric
and rational cases the conditions on $M_m$ become less restrictive
but still these numbers may not be equal to each other.

\subsubsection*{Difference equations for $Q_m$'s}

The functions $Q_m (u)$ obey certain linear difference equations.
In principal, they can be obtained from the system of
linear problems (\ref{B26}), (\ref{B27}) at levels $m=1, 2, \ldots k$
by excluding all functions $F$ except those on the boundaries of
the array (\ref{diag}). For $k=2$ this can be done without any problem.
Indeed, in this case there are only two non-trivial linear
equations:
\begin{eqnarray}
&&T^{1}_{s+1}(u)Q_{1}(u-s)
-T^{1}_{s}(u+1)Q_{1}(u-s-2)
=\phi (u-s)\bar Q_{1}(u+s+1), \nonumber \\
&&T^{1}_{s+1}(u+1)\bar Q_{1}(u+s)
-T^{1}_{s}(u)\bar Q_{1}(u+s+2)
=\phi (u+s+3) Q_{1}(u-s-1).
\end{eqnarray}
Dividing both sides of the first equation by $\phi (u-s)$ and
making use of the fact that the r.h.s. does not depend on $u-s$,
one arrives at
\beq
\phi (u-s)Q_1 (u-s+2)+\phi (u-s+2)Q_1 (u-s-2)=
A(u)Q_1 (u-s),
\eeq
where
$$
A(u)= \frac{
\phi (u-s)T_{s-1}^{1}(u+2)+\phi (u-s+2)T_{s+1}^{1}(u)}
{T_{s}^{1}(u+1)}.
$$
Recall that the b.c. is $T_{-1}^{1}(u)=0$. Whence
one obtains famous Baxter's relation
\beq
\phi (u)Q_1 (u+2)+\phi (u+2) Q_1 (u-2)=T_{1}^{1}(u)Q_1 (u)
\label{bax}
\eeq
which is a 2-nd order difference equation for $Q_1$. The second linear
equation yields the same result (recall that $\bar Q_{1}(u)=Q_{1}(u+1)$).

In general the procedure becomes quite involved.
Here we only quote the results:
\beq
\sum_{a=0}^{k}(-1)^{a}T^{a}_{1}(u-a+1)Q_{1}(u-2a+2)=0\,,
\label{B33}
\eeq
\beq
\sum_{a=0}^{k}(-1)^{a}
\frac{T^{a}_{1}(u+a+1)}{\phi (u+2a+2)}
\frac{ Q_{k-1}(u+2a)}{\phi (u+2a)}=0\,.
\label{B34}
\eeq
The equations for $Q_m$'s with $2\leq m \leq k-2$ have a more
complicated form.

\subsubsection*{Factorization formulas}

At last, we are to identify our $Q_m$'s with $Q_m$'s from the usual
nested Bethe ansatz solution. This is achieved by factorization of
the difference operators in (\ref{B33})
and (\ref{B34}) in terms of $Q_m (u)$.
One can prove the following factorization formulas (looking like
discrete Miura transformation):
\begin{eqnarray}
&&\sum_{a=0}^{k}(-1)^{a-k}
\frac{T^{a}_{1}(u-a+1)}{\phi (u+2)} e^{-2a\p _{u}}\nonumber \\
&=&\left ( e^{-2\p _{u}}-\frac{Q_k (u)Q_{k-1}(u+2)}
{Q_k (u+2)Q_{k-1}(u)}\right ) \ldots
\left (e^{-2\p _{u}}-
\frac{Q_{2}(u)Q_{1}(u+2)}{Q_{2}(u+2)Q_{1}(u)} \right )
\left (e^{-2\p _{u}}-
\frac{Q_1(u)}{Q_1(u+2)}
\right )
\label{B35}
\end{eqnarray}
\begin{eqnarray}
&&\sum_{a=0}^{k}(-1)^{a-k}
\frac{T^{a}_{1}(u+a+1)}{\phi (u+2a+2)} e^{2a\p _{u}}\nonumber \\
&=&\left (e^{2\p _{u}}-
\frac{Q_1(u)}{Q_1(u+2)}
\right )
\left (e^{2\p _{u}}-
\frac{Q_{2}(u)Q_{1}(u+2)}{Q_{2}(u+2)Q_{1}(u)} \right ) \ldots
\left ( e^{2\p _{u}}-\frac{Q_k (u)Q_{k-1}(u+2)}
{Q_k (u+2)Q_{k-1}(u)}\right ).
\label{B36}
\end{eqnarray}
Note that these operators are adjoint to each other. The factors
in the r.h.s. resemble $M$-operators (\ref{Z1}).

Let us note that eq.\,(\ref{B36}) can be rewritten as follows:
\begin{eqnarray}
&&\sum_{a=0}^{k}(-1)^{a}
\frac{T^{a}_{1}(u+a-1)}{\phi (u+2a-2)} e^{2a\p _{u}}\nonumber \\
\!\!&\!\!=\!\!&\!\!\left (\!1\!-\!
\frac{Q_1(u+2)}{Q_1(u)}e^{2\p _{u}}
\right ) \!
\left (\!1\!-\!
\frac{Q_{2}(u+2)Q_{1}(u-2)}{Q_{2}(u)Q_{1}(u)}
e^{2\p _{u}} \right ) \!\ldots  \!
\left (\!1\!-\!\frac{Q_k (u+2)Q_{k-1}(u-2)}
{Q_k (u)Q_{k-1}(u)}e^{2\p _{u}}\right )
\label{B37}
\end{eqnarray}
Now the factors in the r.h.s. resemble $M$-operators (\ref{Z2}).
It is possible to show that coefficients of the operator
inverse to (\ref{B37}) give $T_{s}^{1}(u)$:
\begin{eqnarray}
&&\sum_{s=0}^{\infty}
\frac{T^{1}_{s}(u+s-1)}{\phi (u)} e^{2s\p _{u}}\nonumber \\
\!\!\!&\!\!=\!\!&
\!\!\!\left (\!1\!-\!\frac{Q_k (u+2)Q_{k-1}(u-2)}
{Q_k (u)Q_{k-1}(u)}e^{2\p _{u}}\right )^{-1} \!\!\!
\ldots  \!
\left (\!1\!-\!
\frac{Q_{2}(u+2)Q_{1}(u-2)}{Q_{2}(u)Q_{1}(u)}
e^{2\p _{u}} \right )^{-1}\!\!\!
\left (\!1\!-\!\frac{Q_1(u+2)}{Q_1(u)}e^{2\p _{u}}
\right )^{-1}
\label{B38}
\end{eqnarray}

These formulas yield
$T_{1}^{a}(u)$,
$T_{s}^{1}(u)$
in terms of
elliptic polynomials $Q_m$ with roots constrained by the nested Bethe
ansatz equations which ensure cancellation of poles in $T_{1}^{a}(u)$.
The transfer matrices
$T_{s}^{a}(u)$ for $a,s>1$ can be then found with the
help of determinant formulas (\ref{det3}), (\ref{det4}).

\subsection*{Comments and references}

1. With the b.c. (\ref{B2}) HBDE (\ref{A24a}) is known as the
bilinear form of the discrete two-dimensional Toda molecule equation
\cite{Hirota6} (the Toda lattice with open boundaries), an
integrable discretization of the conformal Toda field theory \cite{cToda}.
In particular, at $k=2$ we have a discrete analogue of the Liouville
equation.

The b.c. in $s$ of the form (\ref{B3}) is known in the classical
theory, too. The b.c. of this kind is a hall-mark of {\it forced
("semi-infinite") hierarchies} of non-linear integrable equations
emerging naturally in matrix models of 2D gravity
(see e.g. \cite{KMMOZ}). For instance, in the forced 2DTL hierarchy,
the $\tau$-function $\tau _{n}(t_1 , t_2 , \ldots |\, \bar t_1 ,
\bar t_2 ,\ldots )$ is equal to zero at $n=-1$: $\tau _{-1}=0$
for any $t_i , \bar t_i$.

Similarly, imposing
the analytic condition on the $\tau$-function
of the form (\ref{B5}) is a familiar story
in classical nonlinear integrable equations since the
paper \cite{AKM}, where the elliptic solutions to the KdV equation
were studied. A systematic approach to elliptic solutions of the KP
equation based on the finite-gap integration methods was developed by
I.Krichever \cite{kr1}.

So we see that each one of the boundary and analytic conditions has been
known in the classical theory. However, they never met alltogether.
The specifics of solutions relevant to quantum problems is perhaps
just in the {\it combination} of the above conditions which looks
quite unusual for the soliton theory.

2. The RS model (in continuous time) was introduced in the
paper \cite{RS} as a relativistic generalization of the Calogero-Moser
system of particles. This model was shown to be integrable and the Lax
representation was found. Recently, it was shown \cite{kz} that the
RS system describes the dynamics of poles of elliptic solutions
(zeros of $\tau$-function) to the 2DTL. Equations of motions
for the discrete time analogue of the RS model were written down
in ref.\,\cite{NRK}, where an ansatz for the Lax pair was suggested.
The
close connection with the nested Bethe ansatz was observed and
explained in ref.\,\cite{KLWZ}.

3. The fact that in the elliptic case
degree of the elliptic polynomial $Q_m (u)$
is equal to $M_m =(N/k)m$ (provided $\eta$ is
incommensurable with the lattice spanned by
the two complex periods $1, \tau$ and
$N$ is divisible by $k$)
follows directly from Bethe equations (\ref{NBE}).
Indeed, the elliptic polynomial form of $Q_m (u)$ implies that
if $u_{j}^{m}$ is a zero of $Q_m (u)$,
i.e., $Q_m (u_{j}^{m})=0$, then $u_{j}^{m} +n_1 +
n_2 \tau$ for all integers $n_1 , n_2$ are its zeros too.
Taking into account the well known
monodromy properties of the $\theta$-function, one concludes that this
is possible if and only if
\beq
M_{m+1}+M_{m-1}=2M_{m}\,,
\label{MMM}
\eeq
which has a unique solution
$
M_{m}=(N/k)m$
satisfying the required b.c. This means that the nested
Bethe ansatz scheme
for elliptic $A_{k-1}$-type models is
consistent only if $N$ is divisible by $k$.

In trigonometric and
rational cases the conditions on degrees of $Q_m$'s become
less restrictive since some of the roots can be located at
infinity. The equality in the formula
for $M_m$ becomes an inequality:
$M_m \leq (N/k)m$. A more detailed analysis \cite{Kirillov} shows that
the following inequalities also hold: $2M_1 \leq M_2$,
$2M_2 \leq M_1 +M_3$, $\ldots $, $2M_m \leq M_{m-1}+M_{m+1}$,
$\ldots $, $N=M_k \geq 2M_{k-1}-M_{k-2}$.

4. It should be noted that the family of commuting
transfer matrices generally does not define a quantum system
uniquely. To do that, one should choose a hamiltonian, i.e. take
a particular representative of the commuting family of operators.
So, to introduce a quantum integrable model to be solved,
one has to give a prescription how to get hamiltonian from the
transfer matrix. To be definite, let as consider the $A_1$-case. The
hamiltonians are taken to be linear in $\log T_s $, to wit
\beq
\label{H}
H=\sum _{p}\alpha _{p}\log T_s (u^{(p)}),
\eeq
where $u^{(p)}$ are some particular values of the spectral
parameter. (For example, the hamiltonian of the spin-1/2 Heisenberg
magnet is $H=\p _{u} \log T_1 (u)$ at $u=0$.) If the hamiltonian is
local, one can define the elementary energy function $\varepsilon (u)$
end represent the total energy as $E=\sum _{j} \varepsilon (u_j )$,
where the momenta $u_j$ of "quasiparticles" satisfy Bethe equations.
This very important characteristics of the quantum system depends
on the choice of $H$.

Is there a room for $\varepsilon (u)$ on the classical side?
The answer seems to be in the affirmative. This function might
encode a way in which potentials or fields entering soliton
equations are expressed through the $\tau$-function of a given
hierarchy. (In the most popular example of KdV this is
$U(x)=2\p ^{2}_{x}\log \tau (x)$.) Since $T_s (u)$ has been
identified with a $\tau$-function, eq.\,(\ref{H}) resembles such
expressions in the hierarchies of discrete soliton equations.
The hamoltonians $H$ might then be identified with the corersponding
potentials (or rather logarithms of potentials in the discrete case).

5. The l.h.s. of eq.\,(\ref{B37})
is known as the generating function
for $T_{1}^{a}(u)$.
These formulas for the generating function
coincide with the ones known in the literature
(see e.g. \cite{BR}, \cite{FR}, \cite{KunSuz}). They may be considered
as non-commutative analogues of generating functions for symmetric
multivariable polynomials (see Remark 3 to Sect.\,2).

\section*{Acknowledgements}

I am grateful to I.Krichever, O.Lipan and P.Wiegmann for
collaboration in \cite{KLWZ} and to
K.Hasegawa, S.Kharchev, S.Khoroshkin,
A.Marshakov,
A.Mironov and A.Orlov
for illuminating discussions.
I would like to thank M.Martellini and A.Gamba for hospitality during
the INTAS School "Advances in Quantum Field Theory and Statistical
Mechanics" (Como, Italy, September 1996), where these lectures were
presented.
I also thank
T.Sultanov for help in preparing the manuscript.
This work was supported
in part by grant CEE-INTAS 93-24-94, by RFFR grant 96-02-19085 and
by ISTC grant 015.

\end{document}